%% file: rapport.tex
\documentclass[utf8,final]{stageM2R} 
\usepackage{amsmath,amsfonts,amssymb,amsthm}
\usepackage{stmaryrd}
\usepackage{lscape}
\usepackage{tikz}
\usetikzlibrary{matrix,decorations.pathreplacing, calc, positioning,fit}
\usepackage{colortbl}

\usepackage [linesnumbered,onelanguage,algoruled]{algorithm2e}

\usepackage{appendix}

\usepackage{geometry}
\newtheorem{definition}{Definition}
\newtheorem{notation}{Notation}
\newtheorem{lemma}{Lemma}

\newtheorem{proposal}{Proposal}

\usepackage[
   backend=biber,        
   sorting=none,          
   isbn=false,
   firstinits=true
]{biblatex}
\author{Hugo Mirault}
\supervisors{Pascal Giorgi\\Fabien Laguillaumie}
\location{LIRMM UM5506 - CNRS, Université de Montpellier
}
\title{Distributed and secure linear algebra} 
\track{Master ALGO}  
\abstractfr{
Cryptography is the discipline that allows securing of the exchange of information. In this internship, we will focus on a certain branch of this discipline, secure computation in a network. The main goal of this internship, illustrated in this report, is to adapt a roster of protocols intended to do linear algebra. We want to adapt them to do algebra for matrices with polynomial coefficients.
We then wish to make a complete analysis of the different complexities of these protocols.
}

\begin{document}   
\selectlanguage{english} 
\frontmatter  
\maketitle    

\cleardoublepage  
\tableofcontents 
\mainmatter  


\input{0-intro}

\input{1-Modele}

\input{2_Determinant_d_une_matrice_polynomiale}
\input{3-Conclusion}

\printbibliography[title=Bibliography ]

\appendix
\appendixpage

\input{Complexite_annexe}

\end{document}

%% file: 0-intro.tex
\chapter{Introduction}

\input{0-Intro/0-Intro}

%% file: 0-Intro/0-Intro.tex
Secure multi-party computation (\textit{multi-party computation} or MPC) is a classic problem in the world of cryptography and distributed computing.

In this problem, several players have a secret share of a certain piece of data. The latter is private and represented by all the shares. Whereas individually, the shares held by the players are only random.\\
We then want to find protocols so that players can compute functions on this data without having to publicly reveal information.\\
We, therefore, want assurances of the security of our data.
We want to protect ourselves from potential attackers who would like to take advantage of the protocols to learn about shared data. In our case, we want attackers who are not limited in their computing power to not be able to learn additional information about the shared data.\\

The multiparty computation is relatively new. It was introduced in the two-player scenario in the 80s by A.Yao \cite{Yao1982ProtocolsFS} with the millionaire problem.\\

This problem discusses two millionaires, Alice and Bob, who wish to compare their fortunes. However, each refuses to reveal the amount of their fortune to the other player. The objective then becomes to compute a bit, allowing us to decide which of the two has the greatest fortune. While maintaining the security of their data.\\
Several solutions have been proposed to solve this problem, in particular by Hsiao-Ying Lin and Wen-Guey Tzeng \cite{Millionaires} based on a computation of a set intersection shared between players.\\

A.Yao \cite{Yao1982ProtocolsFS} therefore first introduced a series for the two-player case.\\
J.Kilian \cite{Kilian1988FoundingCO} then showed that any Boolean formula could be computed within the framework of the MPC, with two players, in a constant number of communication turns.\\
D.Beaver, S.Micali, or S.Rogaway \cite{Beaver1990TheRC} gave, under a certain assumption of complexity, protocols with a constant number of \textit{rounds}, for any function computable in polynomial time. However, in practice, these protocols are unusable because of their high complexity.\\
 U. Feige et al. \cite{Feige2002AMM} as well as Y.Ishai, E.Kushilevitz \cite{Ishai1997PrivateSM}, and D.Beaver \cite{Beaver2000MinimalLatencySF} have shown that it is possible to produce MPC protocols, for all problems of the set $NL$\footnote{corresponding to the set of solvable decision problems by a non-deterministic Turing machine, in space bounded by a logarithmic function. $NL \subseteq P$, but we don't know if NL = P}.\\
The problem remains the difficulty of optimizing the protocols, which are not very applicable in real life, because they are too expensive.\\

Today, the main application of MPC remains machine learning. The data on which the models are trained can be shared between different servers, which can, thanks to the protocols, perform learning on these shares without learning information about the data.

There are other applications in decentralized voting to allow a certain number of participants in an election that preserves the confidentiality of their vote and ensures that no player has been able to cheat.
\subsection{Contribution}
Cramer and Damgard \cite{Cramer2001SecureDL} provided protocols for performing linear algebra on coefficient matrices in a finite field.

We now wish to adapt the ideas of their protocols to the case of matrices with polynomial coefficients. We illustrate the object on which we are going to work in figure 1.1.

We, therefore, propose a series of protocols for manipulating secret shares of polynomial matrices.
We then adapt the protocols of Cramer and Damgard with certain ideas of the protocols of Mohassel and Franklin \cite{efficient_Poly_operation}, to allow the computation of the determinant of a polynomial matrix.

We were able to achieve this adaptation through three different protocols for computing the determinant of a polynomial matrix.

In the second step, we make a complete analysis of the different complexities of all these protocols to determine the most effective adaptation.

Finally, we end up making a comparison of the protocols for computing the determinant, to finally give the most interesting protocol of the three.
\begin{figure}[h]
    \centering
$$A(X) =
\begin{pmatrix}
    a_{1,1}(X)& \dots  & a_{1,j}(X) & \dots & a_{1,n}(X)\\
    \vdots   &  & \vdots  &  & \vdots\\
    a_{i,1}(X) & \dots  & a_{i,j}(X) & \dots & a_{i,n}(X)\\
    \vdots   &  & \vdots  &  & \vdots\\
    a_{n,1}(X) & \dots  & a_{n,j}(X) & \dots & a_{n,n}(X)\\
\end{pmatrix}
$$
   \caption{Example of polynomial matrix}
\end{figure}

%% file: 1-Modele.tex
\chapter{Secure Multi-Party Computing and First Protocols}
In this section, we define notions related to Multi-Party Computation. These different notions are useful for understanding the formalization of the problem, its issues, and its difficulties.
\input{1-Modele/1-Modele}
\input{1-Modele/4-Partage_additif}

%% file: 1-Modele/1-Modele.tex
\section{Secure Distributed Computing}

\subsection{Scenario}
The objective of an MPC scenario is to evaluate a publicly known function $F$ by participants $P_1,P_2,...,P_N$.
Each player $P_i$ has a share of a data $x$ kept secret. The objective of the protocol is therefore to evaluate $F(x)$ from the secret shares without revealing any information either about $x$ or its secret shares.\\
The data $x$ is shared between the participants according to a certain sharing scheme known by the players. We'll see an example of such a scheme later in this section.

\begin{notation}
Let $x$ be information defined on any field or ring (Ex: $\mathbb{Q},\mathbb{F}_q$ or $\mathbb{Z}$).\\
We use the notation $[x]$ to designate a share of the data $x$. We use the notation $[x]_i$ to designate, in particular, the share of $x$ held by the player $P_i$
\end{notation}

Thus, the notation $[f(x)]_i$ represents a secret share of the information of $f(x)$. To compute this share from $[x]_i$ , the player $P_i$ must execute a potentially interactive protocol, allowing him to evaluate the function from his shares and that of the other participants. This principle is illustrated in Figure 2.1.

\begin{notation}
We define $K$ as a finite field of size $q$.\\
We define $N$ as the number of players participating in a protocol.\\
$P_i: [\alpha]_i$ means player $P_i$ owns the share $i$ of the data $\alpha$.\\
By convention, in protocols, interactive protocols (which engage in communication between players) are written in bold, while non-interactive protocols are written in italics.

\end{notation}

\input{1-Modele/tikz}

\subsection{Attackers}
In some studies, assumptions are made about the abilities of potential attackers. A distinction is generally made between two possible attacker scenarios:
\begin{itemize}
\item In the first case, it is assumed that an attacker can see all the internal data of a certain number of players (according to information-sharing models, this number is generally bounded by $N/2 $ or $N-1$). The problem then becomes computing the correct evaluation of the function $F$, without the attacker being able to find information on the secret inputs. We usually talk about \textit{honest, but curious, } attackers or \textit{passive} attackers.
\item In the second case, it is assumed that an attacker can then ``corrupt`` a certain number of players, generally limited to a third of the total number of players.\\
The attacker can thus read the internal data of the players he has ``corrupted``, but he can also modify their actions during the protocols. The problem then becomes the correct evaluation of the $F$ function despite the potential erroneous information, the identification of ``corrupt`` players, and of course the security of the data of ``healthy`` players. We then speak of an \textit{active} or \textit{Byzantine} attacker.
\end{itemize}

\subsection{Communication}

To communicate with each other, players have a broadcast channel available. That is to say, a player can send a piece of data on his channel to allow all the other players to see it.

\section{Measures of complexity}

To compare the efficiency of the different protocols that are produced, several complexity measures are defined.

\begin{definition}[Round complexity]
We define the complexity in \textit{round} of a protocol as the number of communication steps necessary for the proper implementation of the protocol.\\
A communication step represents a phase during which each player can send a message in the communication channel and wait for a response. This is equivalent to one \textit{game turn}.
\end{definition}
\begin{definition}[Communication complexity]
We define the communication complexity of a protocol as the number of message bits sent by a player, necessary for the proper implementation of the protocol. We express the complexity in a number of bits necessary to express a data over the finite field $K$ of size $q$. It is considered that the broadcast of a data item $x \in K$ requires only $\log(q)$ communication bits.
\end{definition}
\begin{definition}[Internal complexity]
We define the internal complexity of a protocol as the number of arithmetic operations in the field $K$, necessary per player to carry out the protocol.
\end{definition}

 Thanks to the models and the different measures of complexity, we now have more elements to talk about the MPC. Before discussing the different protocols, it is worth talking about the different secret sharing schemes that exist and are in use.

%% file: 1-Modele/tikz.tex
\begin{figure}
    \centering
\begin{tikzpicture}
 \node[rectangle,draw,
 minimum width = 10cm, 
minimum height = 0.75cm] (r) at (0,0) {Protocol $f$ };
 \draw [dashed, ->](0,3.7) -- (-4,1.6);
 \draw [dashed, ->](0,3.7) -- (-2,1.6);
 \draw [dashed, ->](0,3.7) -- (-1,1.6);
 \draw [dashed, ->](0,3.7) -- (0,1.6);
 \draw [dashed, ->](0,3.7) -- (1,1.6);
 \draw [dashed, ->](0,3.7) -- (2,1.6);
 \draw [dashed, ->](0,3.7) -- (4,1.6);
 
 \node [draw] at (0,4) {$\alpha \in K$};
 \draw [loosely dotted, line width=0.4mm](-1,1.3) -- (2.9,1.3);
 
 \node at (-2.8,3) {\textit{Sharing}};
 \node at (-4,1.3) {$P_1 : [\alpha]_1$};
 \node at (-2,1.3) {$P_2 : [\alpha]_2$};
 \node at (4,01.3) {$P_N : [\alpha]_N$};
 
 \draw [-stealth](-4,1.1) -- (-4,0.4);
 \draw [-stealth](-2,1.1) -- (-2,0.4);
 \draw [-stealth](4,1.1) -- (4,0.4);

  \draw [-stealth](-4,-0.4) -- (-4,-1);
 \draw [-stealth](-2,-0.4) -- (-2,-1);
 \draw [-stealth](4,-0.4) -- (4,-1);
  \node at (-4,-1.2) {$P_1 : [f(\alpha)]_1$};
 \node at (-2,-1.2) {$P_2 : [f(\alpha)]_2$};
 \node at (4,-1.2) {$P_N : [f(\alpha)]_N$};
  \draw [loosely dotted, line width=0.4mm](-1,-1.2) -- (2.9,-1.2);
 
\end{tikzpicture}\\
\caption{Multiparty Computation Protocol Scheme}

    \label{Multiparty Computation Protocol Scheme}
\end{figure}
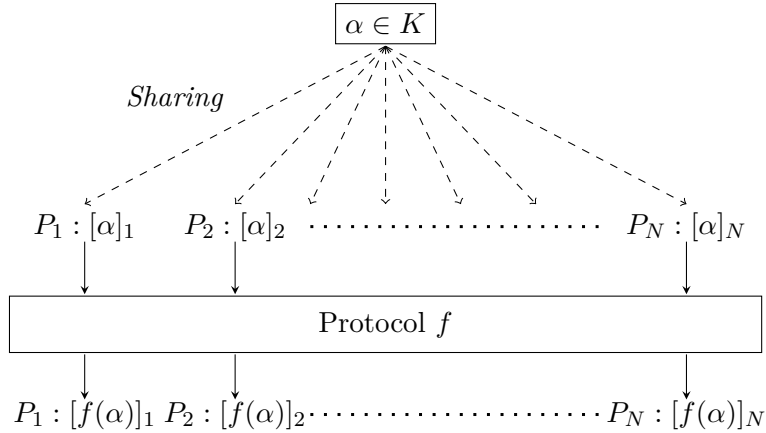

%% file: 1-Modele/4-Partage_additif.tex
\section{Secret sharing schemes}
There are several secret sharing schemes that can handle attackers of different types and with different properties. Historically, Shamir's \cite{shamir} sharing scheme was used to study MPC protocols. This scheme consists in giving each player $P_i$ the evaluation of a polynomial $q_\alpha$ of degree $t$ whose coefficients are random data on $K$ and the coefficient constant is the data $\alpha$ that we want to share. Thereby, it is necessary to obtain the cooperation of $t+1$ players to reconstruct the secret, thanks to the uniqueness of the Lagrange interpolation.\\

More generally, whatever the sharing scheme is, we want to have several properties in order to be able to manipulate these secret shares.
\begin{itemize}
    \item Players can generate random shared data without interaction.
    \item Players can reveal and rebuild their secret data over a constant number of rounds.
    \item Players can add shares, multiply a share by public data, without interaction ($[a]+[b]=[a+b] \quad [a]\cdot c = [a\cdot c]$ ).
    \item Players can multiply two shared secrets in constant rounds.
\end{itemize}

Shamir secret sharing scheme has these properties. The only operation causing a problem is the interactive multiplication of two secrets in a constant number of rounds (see the BGW method \cite{methodeBGW}). Nevertheless, the high cost of these operations leads us to turn to a second secret sharing scheme.\\

We therefore introduce the additive secret sharing of data $\alpha \in K$ between $N$ players. With this share of secrets, the protocols can afford a number $t = N-1$ of corruptible players by a \textit{passive} type of attacker.
To do this, it is assumed that a trusted third party was able to achieve this sharing before the start of any protocol.
Sharing is done as follows:\\

\begin{itemize}
    \item The trusted third party uniformly and independently draws $N-1$ integers out of $K$, denote them $\alpha_i$ for $i$ from $1$ to $N-1$.
    \item The trusted third party compute $\alpha_N = \alpha - (\alpha_1 + \alpha_2 + \cdots + \alpha_{N-1})$
    \item The trusted third party communicates to each player $P_i$ the data $\alpha_i = [\alpha]_i$ which then becomes his secret share.
\end{itemize}

Thus, each player $P_i$ is aware of a secret sharing $[\alpha]_i$ of $\alpha$. The sum of all the secrets makes it possible to reconstitute $\alpha$.\\
We also note that $N-1$ secret data does not allow to reconstitute information on $\alpha$ as long as $\alpha_i$ are uniform and random on $K$.\\

More generally, it is considered a data on a finite ring is shared between players when the latter possess a random element of this ring such that the sum of the secrets makes it possible to reconstitute the data. Little importance is given to how the secret was shared upstream. \\

\section{Basic arithmetic operations}
We will see in this section that the additive secret sharing meets our expectations in terms of the properties listed previously.\\

\textit{Generating a random element: }To generate a split of a random element of a finite field $K$, players individually generate a random, uniform element of $K$, then treat it as the split secret. Knowing that the sum of random and uniform elements over a finite field is a random element over this same field, we can affirm that the generation of the secret shares of a random element is not a problem, assuming that the players can randomly drag an element into a finite field.\\

\textit{Reconstruction of the secret: }To reconstruct a secret shared between the players, the participation of the $N$ players is necessary. Each player sends his private data in clear on the communication network. After knowing the data of the other players, an addition of $N$ elements makes it possible to reconstitute the data. The round complexity is therefore constant. We note \textbf{Reveal}() this interactive protocol.\\

\textit{Adding two secrets: }Players can simply add their secrets.
$$a+b = [a]_1+\cdots +[a]_N + [b]_1 + \cdots [b]_N = \sum_{i = 1}^N[a]_i+[b]_i =\sum_ {i=1}^N[a+b]_i$$

\textit{Multiplication of a secret and a public data: }In the same way, to obtain the secret share, from the multiplication of a public data by a secret share, the classic operation is performed internally.

$$a\cdot c =([a]_1+\cdots +[a]_N ) \cdot c =([a]_1\cdot c+\cdots +[a]_N\cdot c ) = \sum_{i = 1}^N[a]_i\cdot c = \sum_{i = 1}^N[a\cdot c]_i$$
\subsection{Multiplication of two shared secrets}
    We consider that the $N$ players have agreed on a finite field $K$, possess a secret sharing of an integer $a$ and an integer $b$ over the field $K$.\\
    Here we have a less trivial problem to deal with than the previous ones. Indeed, the multiplication of $[a]_i$ and $[b]_i$ by the player $P_i$ is not sufficient, because
    
    $$a \cdot b \neq ([a]_1[b]_1)+\cdots +([a]_N[b]_N) $$
    $$a \cdot b = \sum^N_{i=1} [a]_i \cdot \sum^N_{i=1} [b]_i = ([a]_1 + \cdots + [a]_n) \cdot ([b]_0+\cdots + [b]_N)$$.\\
    
    For this problem, Beaver \cite{crypto-1991-1013} offers a solution called Beaver's triple method. We suppose that, by an external method or thanks to an oracle, the players possess shared secrets of $[x],[y],[z]$ on $K$, such that $x = y *z$ with $ uniform and random y$ and $z$ on $K$. We will discuss this generation later in this report.\\
    From these triples, we can perform an interactive multiplication of two shared data.

\begin{algorithm}[H]
\SetAlgoLined
\SetKw{KwBy}{by}
\KwData{
$N$ players $P_1,\cdots,P_N$\\
$P_i : [a]_i,[b]_i \in K$\\
$P_i : [x]_i,[y]_i,[z]_i \in K$ such that $x = z\cdot y$
}
\KwResult{
$P_i : [a\cdot b]_i \in K$
}
\vspace{5pt}
$P_i :[d]_i \gets [a]_i - [y]_i$ \\
$P_i :[e]_i \gets [b]_i - [z]_i$ \\
$d \gets \textbf{Reveal}([d]_i)$\\
$e \gets \textbf{Reveal}([e]_i)$\\
$P_i : [a\cdot b]_i \gets [x]_i + [y]_i\cdot e +d \cdot [z]_i + d[e]_i$\\
\textbf{Return }$P_i:[a\cdot b]_i$
\caption{\textbf{Multiplication}$(P_i : [a]_i,[b]_i,[x]_i,[y]_i,[z]_i )$}
\end{algorithm}
   
    \begin{center}
        \textbf{Protocol correction:}\\
    \end{center}
    \begin{proposal}
    Algorithm 1 is correct and computes $[a\cdot b]$ the split of the product $a \cdot b$.
    \end{proposal}
    \begin{proof}
By definition of $d$ and $e$, we know that $a = y + d$ and $b = z + e$. The player $P_i$ has at the end of the protocol the data $[x]_i + [y]_i \cdot e + d \cdot [z]_i + d\cdot [e]_i$.

    \begin{equation*}
    \begin{split}
     & \sum_{i = 1}^N \textbf{ Multiplication}([a]_i,[b]_i,[x]_i,[y]_i,[z]_i)\\
    = & \sum^N_{i = 1}([x]_i + [y]_i \cdot e + [z]_i \cdot d + d\cdot [e]_i )\\
   = & \sum^N_{i = 1}[x]_i + \sum^N_{i = 1}[y]_i \cdot e + \sum^N_{i = 1}[z]_i \cdot d + d\cdot \sum^N_{i=1}[e]_i \\
   = & x + y\cdot e + d \cdot z + d \cdot e\\
   = & y\cdot z + y\cdot e + d \cdot z + d \cdot e\\
   = &(y+d)\cdot (z+e)\\
   = & a \cdot b
    \end{split}
\end{equation*}
Thus, if the $N$ players decide to reconstitute the output of this algorithm, they indeed find the product $a \cdot b$. They therefore have a share of $a\cdot b$.
    \end{proof}
   \begin{proposal}
   Algorithm 1 reveals no information about $a$, $b$ and $a\cdot b$.
   \end{proposal}
   \begin{proof}
   At the beginning of the protocol, the players have their secret data. Thus, the only way to gain information is by learning information from other players. However, the only data exchanged are $d$ and $e$ which are made public.\\
   The variables $d$ and $e$ are nevertheless completely random, knowing that they result from the addition of $a$ and $b$ with random and uniform data on $K$. Players cannot therefore derive any information about $a$, $b$ or $a\cdot b$ with knowledge of $d$ and $e$.
   \end{proof} 
    \begin{center}
        \textbf{Protocol complexities: \\}
    \end{center}
    
    The protocol uses two operations \textbf{Reveal} to reveal $d$ and $e$, these operations are in constant number of rounds, thus the protocol has a constant round complexity.\\
    The protocol involves interactions between players only during Reveal operations. The latter require players to send $d,e \in K$ over the network, i.e. $O(\log(q))$ communication bits. So we have a communication complexity in $O(\log(q))$ bits.\\
    Finally, the internal complexity requires two additions for the first two lines. The reconstruction operations of line 3 and 4, require adding the share of all the other players, that is $N$ addition per reconstruction per player. Finally, line 5 engages three additions and three multiplications. We therefore have an internal complexity of $O(N)$.
      
    $$
 \begin{array}{|c|c|c|c|c|}
 \hline
    \textbf{Round}&
    \textbf{B. Triples on $K$}&
    \textbf{Communication} &
    \textbf{Internal}\\\hline
      O(1) & O(1) & O( \log(q)) & O(N)\\\hline
     
    \end{array}
    $$

\section{Generalization of Beaver triples}
In the rest of this internship, we want to handle different algebraic structures of a finite field $K$. We want to be able to manipulate structures such as matrices or polynomials.
Beaver's protocol of interactive multiplication, of secret sharing of elements of a finite field, can be generalized without modification to any ring possessing the right properties. Thus, from a Beaver triplet in a ring $\mathcal{A}$, we can carry out a multiplication of living secrets in this same ring. The only properties needed are those related to protocol security and correctness. We therefore need the following properties.
\begin{itemize}
    \item Players can perform the addition and multiplication of a secret by a public constant without interaction (correction).
    \item The ring is of finite size, we can choose an element randomly and uniformly on it.
    \item The addition of a data $a$ with a random and uniform element returns a random and uniform element on the ring. We can therefore hide values by addition.
\end{itemize}
By having these properties, players can perform all protocol operations.\\
Thus, one can carry out interactive multiplications in the ring of matrices of dimension $n$ with coefficient in a finite field $K$, $\mathcal{M}_n(K)$. This ring has the correct properties and is therefore ``eligible`` for Beaver's interactive multiplication method.
We then only need matrix Beaver triples, that is to say three matrices $X,Y,Z \in \mathcal{M}_n(K)$ such that $X = Y\cdot Z$. The protocol then remains exactly the same as that described previously. Indeed, we can notice that the correctness of the protocol does not require any commutative property.\\
In the same sense, we want to be able to multiply, for example, polynomials. Nevertheless, the ring $K[X]$ of polynomials with coefficients in $K$, is not of finite size. A value cannot therefore be completely masked by addition. We therefore choose to reveal an upper bound $d$ on the degree of the manipulated polynomials. Thus, we can immerse the manipulated elements in $K[X]_{/X^d}$ (the elements of $K[X]$ quotient by $X^d$) and thus be able to mask by addition $y$ and $z$ (an element of $K[X]_{/X^d}$ added to a random and uniform element of $K[X]_{/X^d}$ is a random and uniform element on $K [X]_{/X^d}$).\\
We conclude that from Beaver triples in a ring $\mathcal{A}$ with the right properties, we can perform the multiplication operation.\\

\begin{definition}[Beaver triple complexity]
We define the Beaver triples complexity as the number of interactive multiplications necessary for the good realization of a protocol. This corresponds to the number of beaver triplets that must be obtained upstream of the protocol.\\
It is necessary to distinguish several types of Beaver triples, relative to the type of multiplication that we want to perform.
The nature of the triples is different if we want to perform a multiplication of elements of $K$, a matrix product on $\mathcal{M}_n(K)$ or a polynomial product on $K[X]$.
\end{definition}

The question then arises of the generation of Beaver triples.
In the context of low-level triples, i.e. triples over a field $K$, we have not found any obvious techniques allowing their computation using additive sharing. \\
Nevertheless, in the literature, triples can be generated with unconditional security under Shamir's secret sharing.
To do this, players generate two shared random elements $y,z$.
Then multiply them during an interactive pre-computation phase with, for example, the BGW protocol \cite{methodeBGW}. \\
In our case, we will consider having an oracle capable of providing us with these particular Beaver triples, we note it $BT_K()$. In general, we will note $BT_\mathcal{A}()$ the protocol which allows us to generate Beaver triples on the ring $\mathcal{A}$.\\

\input{1-Modele/BT_schema}

To generate polynomial Beaver triples $BT_{K[X]_{<d}}()$, or matrix $BT_{\mathcal{M}_n(K)}()$, protocols can be proposed from interactive multiplication, by Mohassel and Franklin \cite{efficient_Poly_operation} for the multiplication of polynomials and by Bar-Ilan and Beaver \cite{BarIlan1989NoncryptographicFC} for the matrix product.
However, these protocols consume classic Beaver triples.\\
The idea is simply to generate two random elements on the ring $\mathcal{A}$ then, then multiply them with the protocols they offer.
We then denote, with the input and the result of the multiplication, $x,y,z \in \mathcal{A}$, a new Beaver triple.\\

These protocols are more expensive, but can also be carried out in advance and then, during classic protocols, multiply elements of $\mathcal{A}$ with the correct Beaver triples. This idea is illustrated with figure 2.2.\\

We will not describe these protocols in this report, but we will propose a protocol for generating Beaver Triplets of polynomial matrices under the same principle and using the idea of Franklin and Mohassel. The complexities of these different generation protocols of Beaver triples are proposed in the appendix. \\

We now have the basic tools to propose MPC protocols. There are other useful protocols that we have not mentioned in this part, such as the inversion of an element in a field or the multiplication of $n$ values in constant round. Nevertheless, these protocols are not necessarily essential for the following and will be cited as needed, and their complexities are available in the appendix. We are therefore going to introduce a series of protocols on polynomial matrices, in order to obtain the tools necessary for the protocols for computing the determinant.

%% file: 1-Modele/BT_schema.tex
\begin{figure}
    \centering

\begin{tikzpicture}

\node [rectangle, draw,
minimum width = 2cm, 
 minimum height = 5cm] 
(r) at (-4.5,0) {$BT_{K[X]}()$};

\node at (-6.5,2.3) {$P_1 $};
\node at (-6.5,-2.3) {$P_N $};
\draw [loosely dotted, line width=0.4mm](-6.5,2.1) -- (-6.5,-2.1);
\draw [-stealth](-6.3,2.3) -- (-5.5,2.3);
\draw [-stealth](-6.3,-2.3) -- (-5.5,-2.3);
\draw [-stealth](-3.5,2.3) -- (-2.9,2.3);
\draw [-stealth](-3.5,-2.3) -- (-2.9,-2.3);
\node at (-1,2.3) {$P_1:\left< [x]_1,[y]_1,[z]_1\right> $};
\node at (-1,-2.3) {$P_N: \left<[x]_N,[y]_N,[z]_N\right> $};
\draw [-stealth](0.7,2.3) -- (3,2.3);
\draw [-stealth](0.9,-2.3) -- (3,-2.3);
\node [rectangle, draw,
minimum width = 2cm, 
 minimum height = 5cm]  at (4.5,0) {$multiplication$};
 \node  at (4.5,0.6) {$Interactive$};
 \draw [-stealth](5.9,2.3) -- (6.7,2.3);
\draw [-stealth](5.9,-2.3) -- (6.7,-2.3);
\node at (8,2.3) {$P_1:[\alpha\cdot \beta]_1 $};
\node at (8,-2.3) {$P_N:[\alpha\cdot \beta]_N $};
\draw [loosely dotted, line width=0.4mm](8,2) -- (8,-2);

\node [draw] at (0.4,0) {$\alpha,\beta$};
\node  at (1.8,0) {$share$};
\draw [dashed, ->](0.4,0.3) -- (2.1,2.3);
\draw [dashed, ->](0.4,-0.3) -- (2.1,-2.3);

\node at (2.1,2.6) {$[\alpha]_1,[\beta]_1 $};
\node at (2.1,-2.6) {$[\alpha]_N,[\beta]_N $};

\node [align=center] at (-4.5,3.2) {Interactive \\pre-computation phase};
\node at (-4.5,1.2) {$BT_{K}()$};
\node at (-4.5,-1.2) {$BT_{\mathcal{M}_n(K)}()$};

\end{tikzpicture}\\
\caption{Beaver triple pre-computation system Scheme}

\end{figure}
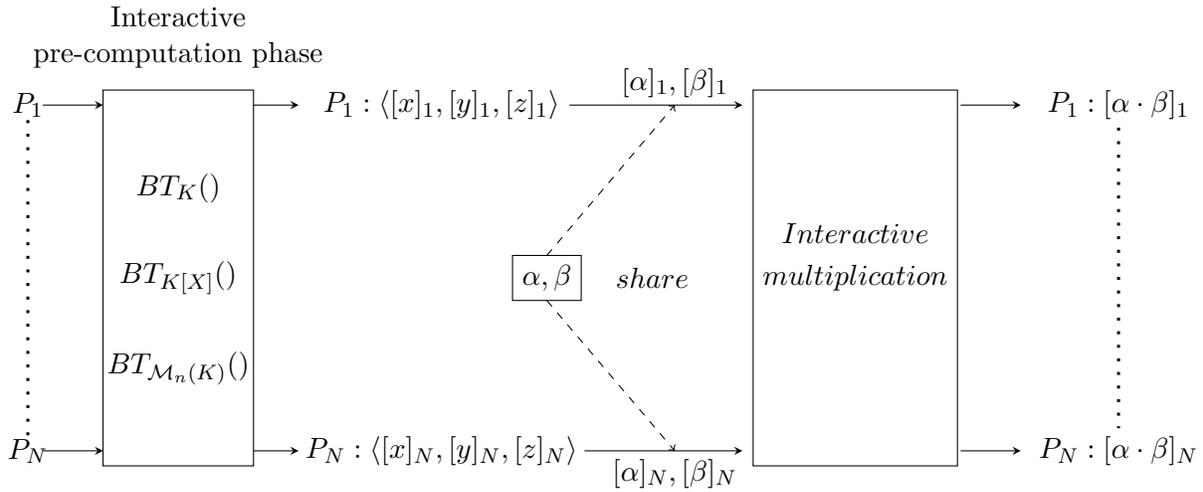

%% file: 2_Determinant_d_une_matrice_polynomiale.tex
\chapter{Basic protocols on polynomial matrices }
We have seen in the previous chapters, MPC protocols allowing us to perform computation on a field $K$. We now turn to the problem posed by the matrices whose coefficients belong to a ring of polynomial $K[X]$.\\
Several protocols have already been proposed to perform operations on a matrix ring by notably Cramer and Damgard \cite{Cramer2001SecureDL}. Other protocols for manipulating polynomials have been proposed by Mohasell and Franklin \cite{efficient_Poly_operation}. Our goal is now to adapt a mixture of these different protocols in order to be able to manipulate the polynomial matrix structure and in a second step, to develop more advanced protocols such as the determinant computation.\\

\begin{definition}
Let $\mathcal{M}_n(K[X]_{<d})$ be the ring of matrices of dimension $n \times n$ with coefficients in the ring of polynomials $K[X]$ of degree at most $d$.
\end{definition}
\begin{definition}
Let $\mathcal{M}_n(K\llbracket X \rrbracket_{/X^{nd+1}})$ be the ring of matrices of dimension $n \times n$ with coefficient in the ring of truncated series $ K\llbracket X \rrbracket_{/X^{nd+1}}$ to $nd+1$. We denote this ring $\mathcal{R}$.
\end{definition}
\begin{definition}
Let $\mathcal{R^*}$ be the subset of invertible matrices of $\mathcal{R}$.
\end{definition}

Let $A(X) \in \mathcal{M}_n(K[X]_{<d})$ be a polynomial matrix which will be shared between the $N$ players during the protocols that will follow. Thus, the players will have a secret share $[A(X)]$ such that $\sum_{i=1}^N[A(X)]_i = A(X)$.\\

It is important to note that we can see $A(X)$ as a matrix with coefficient in the ring $K[X]$ or as a polynomial with coefficient in $\mathcal{M}_n(K)$ . The following example demonstrates this property.
$$A =
\begin{pmatrix}
(2 + 3X + 7X^2) & (8 + 5 X+1X^3)\\
(9 + 4X +6X^2) & (1X^2 +2X^3)\\
\end{pmatrix}
=
\begin{pmatrix}
2 & 8\\
9& 0\\
\end{pmatrix} +
\begin{pmatrix}
 3 & 5\\
4 & 0\\
\end{pmatrix} \cdot X+
\begin{pmatrix}
7& 0\\
6 & 1 \\
\end{pmatrix} \cdot X^2 +
\begin{pmatrix}
0 & 1 \\
0 & 2 \\
\end{pmatrix} \cdot X^3
$$
It is recalled that operations, such as the addition of shares or the multiplication of a secret share with publicly known data, are done internally by applying the operation directly to the secret shares. Let $A(X),B(X),C(X) \in \mathcal{M}_n(K[X]_{<d})$\\
$$[A(X)]+[B(X)] = [A(X) + B(X)]$$
$$[A(X)]\cdot C(X) = [A(X)\cdot C(X)]$$

Nevertheless, these operations have a significant cost from the point of view of arithmetic complexity. It therefore seems necessary to define the notations that represent the product complexities of different structures. The algorithms are in reference \cite{AEPCF},\cite{Bostan2010} and \cite{ModernCA}.

\begin{equation*}
    \begin{split}
        \mathsf{M}(d) &= \text{ Complexity of polynomial multiplication in } K[X]_{<d}\\
             &= O(d^2) \text{ With a naive algorithm}\\
             &= O(d^{\log(3)}) \text{With Karatsuba's algorithm }\\
             &= O(d\log(d)) \text{ By Fast Fourier Transform}\\
             \\
        \mathsf{MM}(n) &= \text{ Complexity of matrix multiplication in } \mathcal{M}_n(K)\\
            &= O(n^{\omega}) \text{Conventional notation with } 2\leq \omega \leq 3\\
             &= O(n^3) \text{ With a naive algorithm}\\
             &= O(n^{\log(7)}) \text{ With Strassen's algorithm}\\
             &= O(n^{2.38}) \text{ By the Coppersmith-Winograd algorithm}\\
             \\
        \mathsf{MM}(n,d) &= \text{ Complexity of polynomial matrix multiplication in } \mathcal{M}_n(K[X])_{<d}\\
             &= O(n^3\mathsf{M}(d)) \text{ With a naive algorithm}\\
             &= O(\mathsf{MM}(n)d\log(d)+ n^2d\log(d)\log\log(d)) \text{ With the Cantor-Kaltofen algorithm \cite{mulmatpolyCantorKaltofen}}\\
             &= O(\mathsf{MM}(n)d + n^2\mathsf{M}(d)) \text{ With the Bostan-Schost algorithm\cite{Bostan2010}}\\
             \\
    \end{split}
\end{equation*}

During several protocols, we will have to mask certain values, in order to reveal them and to apply known algorithms to them. The difficulty is to reveal the masked secret without revealing information on the secret data.
\begin{lemma}
Let $A(X) \in \mathcal{R^*}$ be a fixed polynomial matrix and $R(X) \in \mathcal{R^*}$ a polynomial matrix drawn randomly and uniformly on $\mathcal{R^ *}$. The result of the product $A(X)R(X)$ is uniform and random on $\mathcal{R^*}$.
\end{lemma}
\begin{proof}
Let $A(X)$ and $B(X) \in \mathcal{R^*}$ be two fixed polynomial matrices.
Let $Y$ be a random variable which takes its values uniformly on $\mathcal{R^*}$. We want to show that $Pr(A(X)Y = B(X) ) = \frac{1}{|\mathcal{R^*}|}$.\\
We know that $A(X)$ is invertible in $\mathcal{R^*}$, let $C(X)$ denote the matrix such that $C(X)A(X) = I_n$.\\
Then $Pr(A(X)Y = B(X)) = Pr(Y = C(X)B(X)) = \frac{1}{|\mathcal{R^*}|}$.\\
Thus, the product $A(X)Y$ corresponds to a uniform draw on $\mathcal{R^*}$ of a random variable.
\end{proof}
We therefore know that the product of an invertible secret polynomial matrix, with a uniformly and randomly drawn matrix on $\mathcal{R^*}$, is a randomly drawn matrix on $\mathcal{R^*}$. So we can use this technique to hide and reveal matrices without revealing information about the shared secret.
\input{3-Polynomes_Matriciels/1-Protocoles_basiques}
\input{3-Polynomes_Matriciels/2-Calcul_du_determinant}

%% file: 3-Polynomes_Matriciels/1-Protocoles_basiques.tex
\section{Evaluation of a polynomial matrix}
We want the players to retrieve a secret share of the evaluation of the matrix $A$ at a point $\alpha$.
\begin{lemma}
The evaluation of a partition of a polynomial matrix at a public point is equivalent to the evaluation at a public point.\\
That is, the matrix evaluation operation can be performed internally.
$A(\alpha) = \sum_{i = 0}^N[A]_i(\alpha) $
\end{lemma}
\begin{proof}
 We can represent the matrix $A$ as a polynomial with coefficient in $\mathcal{M}_n$, but also as a matrix with coefficient in $K[X]$.

$$
 A=
\begin{pmatrix}
a_{1,1}(X)&\cdots & a_{n,1}(X)\\
\vdots& \ddots & \vdots \\
a_{1,n}(X)&\cdots & a_{n,n}(X)\\
\end{pmatrix}
= \begin{pmatrix} A_0 \end{pmatrix} + \begin{pmatrix} A_1 \end{pmatrix} \cdot X + \cdots + \begin{pmatrix} A_d \end{pmatrix} \cdot X^d
$$

$A_j \in \mathcal{M}_n(K)$ are the matrix coefficients of the polynomial $A(X)$.
So we have :
 $$A(\alpha) = \sum_{j = 0}^d A_j \cdot \alpha^j = \sum_{i = 0}^N \left[\sum_{j = 0}^d A_j \cdot \alpha^j\right]_i = \sum_{i = 0}^N \sum_{j = 0}^d [A_j]_i \cdot \alpha^j = \sum_{i = 0}^N [A]_i (\alpha) $$
 Thus, the reconstruction of the secret partitions evaluated at a point $\alpha$ makes it possible to find the initial polynomial evaluated at $\alpha$.
\end{proof}

\section{Interpolation of polynomial matrices}

The players own $d+1$ matrices $b_j \in \mathcal{M}_n(K)$, secretly shared and $d+1$ publicly known point $a_j \in K$. We want players to be able to find the unique matrix polynomial of degree at most $d$, $P(X)$ , such that $P(a_j) = b_j $
\begin{lemma}
The interpolation protocol of a matrix polynomial of degree $\leq d$, from $d+1$ secret sharing of matrices and $d+1$ publicly known evaluation points, is feasible without interactions between players.
\end{lemma}
\begin{proof}
 Let $b_0,\cdots,b_d \in \mathcal{M}_n(K)$ and $a_0,\cdots,a_d \in K$, we are looking for the unique matrix polynomial $P \in \mathcal{M}_n (K[X]_{<d})$ such that $P(a_i) = b_i$ for all $i$.\\
 According to Lagrange's interpolation theorem, we can find the polynomial $P$ in the following form:

$$ P(X) = \sum_{i=0}^d b_i \prod_{j = 0,j\neq i}^d \frac{X-a_j}{a_i - a_j} = \sum_{i=0 }^d b_i \cdot l_i $$

Here, we note the interpolation coefficients $l_i = \prod_{j=0,j\neq i}^d \frac{X-a_j}{a_i - a_j}$.
In our case, each player $P_h$ has only $[b_i]_h$ as information on the matrix $b_i$. Nevertheless, the $a_i$ being known publicly, each player can directly compute the $l_i$, we can thus consider that this data is public. By interpolating directly on the secret shares, the player $P_h$ obtains the following data $ \sum_{i=0}^d [b_i]_h l_i$.
$$\sum_{h=1}^N \sum_{i=0}^d [b_i]_h \cdot l_i = \sum_{h=1}^N \sum_{i=0}^d [b_i \cdot l_i ]_h = \sum_{i=0}^d b_i \cdot l_i = P(X) $$

We then observe that if all the players add the interpolation of their sharing, they indeed find the interpolated shared polynomial. The additive sharing scheme is preserved through interpolation at public points.
\end{proof}

\section{Interactive multiplication of two polynomial matrices}
We want to have a protocol such that, $N$ players having the secret sharing of two polynomial matrices $A(X),B(X) \in\mathcal{M}_n(K[X]_{<d} )$, are able to compute the secret share $[A(X)\cdot B(X)]$ of the product of the two matrices.\\

As discussed previously, we will use the Beaver triplet method already seen previously. The players therefore have a triple $X,Y,Z$ such that $X = Y\cdot Z$ and $Y,Z$ are uniform and random on $\mathcal{M}_n(K[X]_{<d })$. We want to multiply the shares of $A(X)$ and $B(X)$.

\begin{itemize}
    \item Players internally compute and reveal $[D] = [A] - [Y]$ and $[E] = [B] - [Z]$.
    \item Players compute the product of the shares
    $$ P_i: [A\cdot B]_i = [(D+Y)(E+Z)]_i = [DE + DZ + YE + YZ]_i = [D]_iE + D[Z]_i + Y[ E]_i + [X]_i$$
\end{itemize}
The complexities naturally arise as we described them in the previous section. The round complexity, and the number of triples on $\mathcal{M}_n(K[X]_{<d})$, are indeed constant. The polynomial matrices $D(X)$ and $E(X)$ which are revealed therefore entail a communication complexity proportional to their size, hence $O(n^2d\log(q))$. And finally for the internal complexity, we have a constant number of multiplications and addition of matrix polynomials($O(\mathsf{MM}(n,d)$), we add the cost of the reconstruction of the two matrices $D (X)$ and $E(X)$, in $O(n^2dN)$.
$$
 \begin{array}{|c|c|c|c|}
 \hline
    \textbf{Round}&
    \textbf{B. Triples on $\mathcal{M}_n(K[X]_{<d})$}&
    \textbf{Communication} &
    \textbf{Internal}\\\hline
      O(1) & O(1) & O(n^2d\cdot \log(q)) & O(\mathsf{MM}(n,d) + n^2dN) \\\hline
     
 \end{array}
$$

We now ask ourselves the question of the generation of our Beaver triples of matrix polynomials. The proposed protocol is executed upstream of the other protocols as an interactive pre-computation, and thus its complexity, higher than the multiplication by the Beaver triples, does not count in the cost of the protocols. Nevertheless, it remains interesting to know its costs and the technique used.\\

To generate our triples, we use the idea proposed by Mohassel and Franklin for the multiplication of two secretly shared polynomials \cite{efficient_Poly_operation}.
We will need a protocol for the interactive multiplication of two matrices on $\mathcal{M}_n(K)$ from Beaver triples on $K$.
This last protocol is proposed by Bar-Ilan and Beaver \cite{BarIlan1989NoncryptographicFC}, we note it \textbf{MulMat}$(P_i: [A]_i,[B]_i) \rightarrow (P_i:[A\cdot B] _i)$.
We will also need the non-interactive polynomial matrix interpolation and evaluation protocols discussed earlier.
Finally, we will need a non-interactive protocol for generating a random and uniform polynomial matrix on $\mathcal{M}_n(K[X]_{<d})$, we note it \textit{RandPolyMat} (n,d).
 
\begin{algorithm}[H]
\SetAlgoLined
\SetKw{KwBy}{by}
\KwData{
$N$ players $P_1,\cdots,P_N$\\
$d$ the desired maximum degree of Beaver triples
}
\KwResult{
$P_i : [A],[B],[C] \in \mathcal{M}_n(K[X])$ with $B,C$ of degree at most $d$ and $A$ of degree at most $2d$, such that $A(X) = B(X)\cdot C(X) $.
}
\vspace{5pt}

$P_i: [B] \gets \textit{RandPolyMat}(n,d)$\\
$P_i: [C] \gets \textit{RandPolyMat}(n,d)$\\
\For{$j\gets 1$ to $2d+1$}
{
$[B(j)]\gets$ \textit{Rating}$([B(X)],j)$\\
$[C(j)]\gets$ \textit{Evaluation}$([C(X)],j)$\\

$P_i: [B(j)\cdot C(j)]_i \gets$ \textbf{MulMat}$(P_i: [B(j)]_i,[C(j)]_i)$\\
}
$P_i: [A(X)]_i \gets$ \textit{InterpolationPolyMat}
$(P_i : [B(1)\cdot C(1)]_i,\cdots ,[B(2d+1)\cdot C(2d+1)]_i,(1,\cdots,2d+1)) $\\
\textbf{Return }{$P_i : ([A(X)]_i,[B(X)]_i,[C(X)]_i)$}
\caption{\textbf{$BT_{\mathcal{M}_n(K[X]_{<d})}$}$(P_i : \emptyset )$}
\end{algorithm}
 
 \begin{proposal}
 The \textbf{$BT_{\mathcal{M}_n(K[X]_{<d})}$} protocol returns the secret share of a triple $A,B,C$ such that $A(X) = B(X) \cdot C(X)$
 \end{proposal}
 \begin{proof}
 From line 5, each player has a secret share of two random and uniform matrix polynomials on $\mathcal{M}_n(K[X]_{<d})$,$[A(j)],[ B(j)]$ evaluated in $2d+1$ different values.\\ These evaluated are partitions of matrices which are multiplied by the \textbf{MulMat}() protocol.
We therefore have $2d+1$ evaluation of a polynomial of degree at most $2d$. Players can therefore interpolate.
Now we know that $A(j) \cdot B(j) = (A\cdot B)(j)$. So the interpolated polynomial which is returned to the output of the protocol corresponds to the secretly shared product $[A(X)\cdot B(X)]$
 \end{proof}

\textbf{Complexity in rounds:}\\
 The only step that engages a communication is step 6. We know a protocol \cite{Cramer2001SecureDL} of matrix multiplication in constant round. Nevertheless, we are in a loop, as each product $A(j)\times B(j)$ is independent, we can launch them all in parallel. Thus, the protocol is in constant round.\\
 
 \textbf{Beaver triplets on $K$:}\\
 We seek to produce Beaver triples of matrix polynomials using only classical triples.
 The only step that engages communications is step 6. Cramer and Damgard's protocol engages $\mathsf{MM}(n)$ or $n^\omega$ Beaver triples on $K$. Let, for the $2d+1$ iterations, a total of $O(dn^\omega)$ triples required.\\
  
  \textbf{Communication complexity:}\\
   The only step that engages communications is step 6. Cramer and Damgard's protocol for matrix multiplication engages $O(n^\omega\log(q))$ bits of communications. There is $O(d)$ running this protocol, so the MultPolyMat protocol requires $O(dn^\omega\log(q))$ communication bits.\\
   
   \textbf{Internal Complexity:}\\
   For line 1 and 2, we generate an element of size $O(n^2d)$.
   For line 4 and 5, the polynomial matrix evaluation algorithm requires $O(d)$ matrix additions and $O(d)$ matrix multiplication by an integer. So $O(dn^2)$ operation by evaluation, knowing that there is $2d+1$ evaluation, we have an internal complexity in $O(d^2n^2)$.\\
   For line 6 we execute $O(d)$ times the Cramer and Damgard protocol which requires $O(n^{\omega}N)$ arithmetic operations. So this line requires $O(dn^{\omega}N)$.\\
   Line 6 engages a polynomial matrix interpolation algorithm. By taking a naive algorithm, we obtain a complexity in $O(d^2n^2)$.\\
   We therefore obtain an overall complexity for this protocol of $O(dn^{\omega}N+ d^2n^2)$ per player.

$$
 \begin{array}{|c|c|c|c|c|}
 \hline
    \textbf{Round}&
    \textbf{B. Triples on $K$}&
    \textbf{Communication} &
    \textbf{Internal}\\\hline
      O(1) & O(dn^\omega) & O(dn^\omega \cdot \log(q)) & O(dn^{\omega}N+d^2n^2)\\\hline
     
 \end{array}
    $$

\section{Generation of a random and uniform element in $GL_n(K\llbracket X\rrbracket_{/X^{d+1}})$}

We assume that each player knows how to generate a random and uniform element in $K$. This operation is called \textit{Rand()}.
Thus, if each player $P_i$ draws a value $[x]_i$ on $K$, then the sum $\sum_{i=1}^N [x]_i = x$ is a random and uniform element on $ K$. This is how players can generate the secret shares of an item on $K$.\\
From there, players can generate $d+1$ secret share and identify them as the coefficients of a polynomial on $K[X]_{<d+1}$. Players can therefore generate the secret shares of a uniform and random polynomial on $K[X]_{<d+1}$. We note this protocol, \textit{RandPoly($d$)}.\\
Continuing in the same direction, players generate $n^2$ uniform and random polynomials on $K[X]_{<d+1}$ and identify them as the coefficients of a polynomial matrix of dimension $n \times n$. So we have a protocol to generate a uniform and random polynomial matrix on $\mathcal{M}_n(K[X]_{<{d+1}})$. We note this last protocol, \textit{RandMatPoly($n,d$)}. We note that finally this last protocol comes down to asking each player to generate $n^2(d+1)$ elements out of $K$ to form a polynomial matrix and to consider the latter as the secret sharing of a uniform polynomial matrix and random on $\mathcal{M}_n(K[X]_{<d+1})$. Note that each of these protocols does not require any interaction.\\

We then want to be able to generate an invertible element in $\mathcal{M}_n(K\llbracket X\rrbracket_{/X^{d+1}})$.
We know that the invertible elements in the formal series are the elements whose constant term is invertible \cite{ModernCA}. Now Cramer and Damgard \cite{Cramer2001SecureDL} propose an interactive protocol which uniformly generates a matrix of dimension $n \times n$ invertible in $K$, we denote \textbf{RandInvMat}(n) the latter. We then propose the following protocol.
 
\begin{algorithm}[H]
\SetAlgoLined
\SetKw{KwBy}{by}
\KwData{
$N$ players $P_1,\cdots,P_N$\\
$\mathcal{M}_n(K\llbracket X\rrbracket_{/X^{d+1}})$, the target set\\
}
\KwResult{
$F(X) \in \mathcal{M}_n^*(K\llbracket X\rrbracket_{/X^{d+1}})$
}
\vspace{5pt}
$P_i :[F(X)]_i \gets$ \textit{RandMatPoly}$(n,d-1)$\\
$P_i :[G]_i \gets$ \textbf{RandInvMat}$(n)$\\
$P_i : [F(X)]_i \gets [F(X)]_i \cdot X + [G]_i$\\
\textbf{Return }{$P_i: [P(X)]_i$}
\caption{\textbf{RandInvPolyMat}$(n,d)$}
\end{algorithm}
 
 The idea of the protocol is simply to generate a matrix polynomial uniformly over $\mathcal{M}_n(K[X]_{<d-1}))$, shift the coefficients by multiplying by $X$, and add a constant term in $\mathcal{M}_n(K)$.\\
\begin{center}
     \textbf{Protocol complexity}
 \end{center}
 the first line therefore requires no interaction, you must generate $O(dn^2)$ elements on $K$. The arithmetic complexity of \textit{RandMatPoly(n,d)} is in $O(dn^2)$.\\
 The second line, \textbf{RandInvMat(n)} is a protocol proposed by Cramer and Damgard in constant round. This protocol does not use Beaver triples on $K$ but does use one on $\mathcal{M}_n(K)$. The communication complexity of this protocol is in $O(n^2\log(q))$ bits, and its internal complexity is in finally $O(n^{\omega} + N\cdot n^2)$.\ \
 Finally, line 3 simply consists of shifting the coefficients of $F(X)$ and adding the matrix $G$ as a constant coefficient of $F(X)$. There are therefore no arithmetic operations to be performed.

$$
 \begin{array}{|c|c|c|c|c|}
 \hline
    \textbf{Round}&
    \textbf{B. Triples on $K$}&
    \textbf{B. Triples on $\mathcal{M}_n(K)$}&
    \textbf{Communication} &
    \textbf{Internal}\\\hline
      O(1) & - & O(1) & O(n^2\cdot \log(q)) & O(n^{\omega}+dn^2 + N\cdot n^2)\\\hline
     
 \end{array}
$$

We now have the protocols allowing us to manipulate the secret shares of polynomial matrices.
We have therefore defined all the necessary tools to allow us to approach linear algebra in polynomial matrices.

%% file: 3-Polynomes_Matriciels/2-Calcul_du_determinant.tex
\chapter{Computation of the determinant of a polynomial matrix}
We have previously presented a suite of tools allowing us to manipulate shared polynomial matrices. We want to address the computation of the shared polynomial matrix determinant. The computation of the determinant makes it possible to know the singularity of a matrix and therefore to move towards, among other things, polynomial matrix inversion, system resolution.\\

By knowing a bound on the maximal degree of $A(X)$, players can guess a bound on the degree of the determinant of $A(X)$.

\begin{lemma}
Let $A(X) \in \mathcal{M}_n(K[X]_{<d})$
The determinant of $A(X)$ is a polynomial $D(X)$ of degree at most $d\cdot n$
\end{lemma}
\begin{proof}
We can deduce a bound on the degree of the determinant $D(X)$ of $A$ with the expression of the Laplace expansion:
$$D(X) = det(A(X)) = \sum_{j=1}^n (-1)^{i+j} A_{i,j}(X) \cdot m_{i,j }(X) \quad \forall i \leq n$$
$A_{i,j}(X)\in K[X]_{<d}$ represents the polynomial in coordinate $i,j$.\\
$m_{i,j}(X)\in K[X]_{<nd}$ represents the determinant of the matrix $A(X)$ in which we have deleted column $i$ and row $j$ .\\
Let $T_d(n)$ be the expression that defines the bound on the degree of the determinant of a polynomial matrix of degree at most $d$ and dimension $n \times n$. We then have with the previous expression:
$$
T_d(n) \leq d \cdot T_d(n-1) \leq d\cdot n
$$
Thus, the determinant of the matrix $A(X)$ has degree at most $dn$.\\
\end{proof}

Cramer and Damgard \cite{Cramer2001SecureDL} propose a protocol allowing to compute the determinant of a classical matrix with coefficient in a finite field. Nevertheless, in the context of polynomial matrices we find ourselves in the ring of coefficient matrices, in the ring of polynomials. To overcome this problem, several solutions are possible. We'll look at three of these solutions and then compare them in terms of their complexities and their likelihood of success.\\
There are nevertheless two protocols proposed by Cramer and Damgard. When the matrix $A(X)$ is known to be invertible, we use a protocol a little faster than the general case, nevertheless this protocol would reveal the determinant equal to zero if we tried to execute this protocol on a matrix not reversible. This is why a second protocol is proposed for the general case.

\section{Interpolation Evaluation Method}
 The first method to compute the determinant could be to recover evaluates of the determinant then to interpolate the polynomial $D(X)$ which represents the determinant of $A(X)$. It is thus necessary that $dn+1 < q $, to have enough interpolation points.\\
 Thus, each player $P_i$ has a share of the matrix $A(X)$, $[A(X)]_i \in \mathcal{R}$.
 \begin{itemize}
     \item Each player evaluates $A$ on $dn+1$ distinct points internally. Each player then has $[A(i)]$ for $j$ from $0$ to $dn$.
     \item Each player computes the share of the determinant of $A(j)$ with the Cramer and Damgrad protocol. Players run this protocol for $A(j)$, $j$ from $0$ to $dn$.\\
     \textbf{DeterminantCramerDamgard}($P_i: [A]_i, A \in \mathcal{M}_n(K)) \rightarrow P_i: [det(A)]_i$
     \item The players then have $[D(j)]$ for $j$ from $0$ to $dn$, we know that $D(X)$ has degree at most $dn$. Players can then interpolate the $D$ polynomial internally knowing that the $dn+1$ points are publicly known.
     \item The players then hold $[D(X)]$, the secret share of the determinant of $A(X)$.
 \end{itemize}

 \begin{algorithm}[H]
\SetAlgoLined
\SetKw{KwBy}{by}
\KwData{
$N$ players $P_1,\cdots,P_N$\\
$[A(X)] \in \mathcal{R}$ a shared polynomial matrix\\
}
\KwResult{$[det(A(X))] \in K[X]$ the split of the determinant of $A(X)$}
\vspace{5pt}
\ForEach{$j \in \llbracket 0,dn \rrbracket$ }
{
$[A(j)] \gets \textit{Rating}([A(X)],j)$\\
$P_i: [det(A(j))]_i \gets \textbf{DeterminantCramerDamgard}(P_i: [A(j)]_i$)\\
}
$[det(A(X))] \gets \textit{Interpolation}([det(A(0))],\cdots ,[det(A(dn))]) $\\
\textbf{Return:}$P_i: [Det(A(X))]_i$
\caption{\textbf{DetMatPoly\_EvalInterpol}$((P_i : [A(X)]_i)_{i\in\llbracket 1,N\rrbracket})$}
\end{algorithm}

 \begin{center}
    \textbf{Proof of correction}
\end{center}

\begin{proposal}
Algorithm 4 returns the secret share of $det(A(X))$, and does not reveal any information about $A(X)$
\end{proposal}
\begin{proof}
 
 The algorithm evaluates the determinant of the matrix $A(X)$, then uses a polynomial interpolation algorithm to reconstruct the polynomial.\\
 To recover the evaluated values of the determinant of $A(X)$, we first use an evaluation algorithm. As seen previously, the evaluation of the polynomial matrix $A(X)$ at a publicly known point $j$ is done directly internally. Each player therefore has a secret partition $[A(j)]$ of the matrix $A(X)$ evaluated in $j$ for $dn+1$ distinct points.\\
The participants then compute the determinant of the matrix $A(j) \in \mathcal{M}_n(K)$ using the Cramer and Damgard protocol \cite{Cramer2001SecureDL}.\\
The players therefore possess the secret shares $[det(A(j))] = [D(j)]$ for $dn+1$ different values of $j$.
We therefore choose to interpolate the polynomial $D(X)$. This last protocol is realizable without interaction as seen in the previous section.\\
Thus, the players have a secret share $D(X)$ at the end of the protocol. $D(X)$ is therefore the unique polynomial of degree at most $dn+1$ such that $D(i) = det(A(i))$. Knowing that the determinant of $A(X)$ has degree at most $dn+1$, $D(X)$ is the determinant of $A(X)$.\\

We want to preserve any leakage of information on $A(j)$, but we do not know if $A(j)$ is invertible, so we must use the protocol proposed by Cramer and Damgard which protects the case of a matrix not reversible. By doing this, we protect the potential leakage of a root of the determinant of $A(X)$. No other protocol is interactive.\\
\end{proof}

\begin{center}
    \textbf{Complexity proof}
\end{center}

\textbf{Round complexity:}\\

The only protocol that engages in communication is on the $3$ line.
The protocol to compute the determinant by Cramer and Damgard which protect the case of the invertible determinant is indeed in constant round. Knowing that each of the executions of this protocol is independent, we can launch the $dn+1$ executions in parallel and thus preserve the complexity in a constant round.
So we have a constant round complexity for our protocol. \\

\textbf{Beaver triples on $K$:}\\
As before, the only protocol that engages communication is on line 3, so we execute $dn+1$ times the Cramer and Damgard protocol which requires $O(n^2)$ Beaver triples per execution. So for the $dn+1$ iterations, we have $O(dn^3)$ Beaver triples on $K$.\\

\textbf{Beaver triples on $\mathcal{M}_n(K)$:}\\
As for the complexity of the Beaver triples on $K$, only line 3 is expensive. The Cramer and Damgard protocol requires $O(n)$ Beaver triples over $\mathcal{M}_n(K)$. So for the $dn+1$ iterations, we need $O(n^2d)$ Beaver triples on $\mathcal{M}_n(K)$.\\

\textbf{Communication complexity:}\\
As with the previous complexities, only line 3 engages communications. Thus, knowing that the protocol of line 3 commits $O(n^3\log(q))$ communication bits per player. We need, for the $dn+1$ iterations, $O(dn^4\log(q))$ communication bits per player.\\

\textbf{Internal complexity:}\\
The evaluation of a matrix polynomial of degree $d$, as seen previously, can be done like a classical evaluation of a polynomial. Let $A_i$ be the $i^{th}$ matrix coefficient of the polynomial $A(X)$, then $[A(j)] = \sum_{i=0}^d [A_i] j^i $. So we have $d$ multiplication of a matrix by a scalar and $d$ additions of matrices, or $O(dn^2)$ for an evaluation. So we have a complexity for the $dn+1$ iterations, $O(d^2n^3)$.\\
The computation of the determinant by the Cramer and Damgard protocol at line 3 requires $O(n^{\omega+1}+n^3N)$ per iteration, i.e. in total $O(dn^{\omega+2} + n^4dN )$.\\
For line 5, the interpolation of a matrix polynomial requires internal complexity $O(n^2d)$.\\
So we finally have the arithmetic complexity of this protocol in $O(n^3d^2 + dn^{\omega+2} + n^4dN)$.\\
$$
\small
 \begin{array}{|c|c|c|c|c|}
 \hline
    \textbf{Round}&
     \textbf{Beaver Triples on $K$}&
    \textbf{Beaver Triples on $\mathcal{M}_n(K)$}&
    \textbf{Communication} &
    \textbf{internal}\\\hline
     O(1) & O(n^3d) & O(n^2d) & O(n^4dlog(q)) & O(n^3d^2+n^{\omega+2}d+ n^4dN) 
     \\\hline
 \end{array}
 $$

\section{Method $\mod X^{nd+1}$}

The second method to compute the determinant would be to directly adapt the Cramer and Damgard protocol used previously. However, for the protocol for computing the determinant in $\mathcal{M}_n(K)$, we need to compute inverses of elements of $K$. But in the context of polynomial matrices, we need to compute the inverses of elements in $K[X]$. To overcome this problem, we place ourselves in the ring of truncated series at $nd+1$ because our determinant has degree at most $nd$. We can then adapt the protocols.

\subsection{RandMatPolyDet: Generation of a random matrix and it's determinant}

In order to compute the determinant, we need a protocol that generates a mask, a random, and uniform $H$ matrix in $\mathcal{R^*}$. The matrix must be invertible, and its determinant must be known and also secretly shared between the players.\\
We propose here the direct adaptation of the Cramer and Damgard protocol.
It is recalled that within the framework of the truncated series, the invertible elements are those whose constant coefficient is invertible.

\begin{itemize}
    \item The players generate $2n$ uniform invertible values in $K[X]_{/X^{nd+1}}$ which we note $[u_i(X)]$ and $[l_i(X)] $ for $1\leq i \leq n$. These values represent the secret share of $u_i(X)$ and $l_i(X)$ for $1\leq i \leq n$. One can check that a value is invertible in the truncated series ring by checking that its constant coefficient is invertible, which is done with the Bar-Ilan and Beaver protocol \cite{BarIlan1989NoncryptographicFC}.\\
    \textbf{RandPolyInv}$(d)\rightarrow P_i:[G(X)] \in GL(K[X]_{X^{d+1}})$
    \item Players compute with the Fan-In Multiplication protocol, the product $[\prod_{i=1}^n u_i(X)l_i(X) \mod X^{nd+1} ]$ that we note $[d_H(X)]$.\\
     \textbf{Fan-In-MulPoly}$(P_i: [G_1(X)]_i,\cdots,[G_n(X)]_i) \rightarrow P_i: [\prod_{j=1}^n G_j(X) ]_i$
    \item Players internally construct the secret sharing of the $U(X)$ matrix and the $L(X)$ matrix.
    $$
    U(X) =
    \begin{bmatrix}
    u_1(X) & r_{1,2}(X) & \cdots &r_{1,n}(X)\\
    0 & \ddots & \ddots& \vdots\\
    0 & 0 &\ddots & r_{n-1,n}(X)\\
    0 & 0 & 0 &u_n(X)
    \end{bmatrix}
    \quad L (X)=
    \begin{bmatrix}
         l_1(X) & 0 & 0 &0\\
     r'_{2,1}(X)& \ddots & 0 & 0\\
     \vdots& \ddots& \ddots & 0\\
     r'_{n,1}(X)& \cdots& r'_{n,n-1}(X)& l_n(X)
    \end{bmatrix}
    $$
    $r_{i,j}(X)$ and $r'_{i,j}(X)$ are drawn uniformly and randomly on $K[X]_{<nd+1}$.
    \item Players perform an interactive matrix multiplication to compute the product $[U(X)\cdot L(X)]$ which we note $[H(X)]$.
    
\end{itemize}
 
 \begin{algorithm}[H]
\SetAlgoLined
\SetKw{KwBy}{by}
\KwData{
$N$ players $P_1,\cdots,P_N$\\
$\mathcal{R^*}$ the target set\\
}
\KwResult{
$[H(X)] \in \mathcal{R^*}$ random and uniform on $\mathcal{H}$ \\
$[d_H(X)] \in K[X]_{/X^{nd+1}}$ the sharing of the determinant of $H(X)$}
\vspace{5pt}
\ForEach{$j \in \llbracket 1,n \rrbracket$ }
{
$P_i: [u_j(X)]_i \gets \textbf{RandPolyInv}(dn)$\\
$P_i: [l_j(X)]_i \gets \textbf{RandPolyInv}(dn)$\\

}
$P_i: [d_H(X)]_i \gets $\textbf{Fan\_In\_MulPoly}$( P_i: [u_1(X)]_i,\cdots,[u_n(X)]_i,[l_1(X) ]_i,\cdots,[l_n(X)]_i) \mod X^{nd+1}$\\
$[L(X)] \gets $\textit{RandomPolyLowerMatrix}$([l_1(X)],\cdots,[l_n(X)])$\\
$[U(X)] \gets $\textit{PolyUpperRandom Matrix}$([u_1(X)],\cdots,[u_n(X)])$\\
$P_i: [H(X)]_i \gets $\textbf{MultiplicationMatricesPoly}$( P_i: [U(X)]_i,[L(X)]_i) \mod X^{nd+1}$\\

\textbf{Return }$P_i : ([H(X)]_i,[d_H]_i)$

\caption{$\textbf{RandMatPolyDet}(n,nd)$}
\end{algorithm}

\begin{center}
    \large\textbf{Proof of correction}
\end{center}

\begin{proposal}
$H(X) \in \mathcal{R^*}$ and its determinant is $d_H(X)$.
\end{proposal}
\begin{proof}
The values $u_i(X)$ and $l_i(X)$ are drawn from the subgroup of invertible of $K[X]_{/X^{nd+1}}$. As the matrices $U(X)$ and $L(X)$ are triangular matrices, whose diagonal is composed exclusively of invertible elements, we deduce that $U(X)$ and $L(X)$ are invertible and that their determinants are respectively $\prod_{j=0}^n u_j(X) \mod X^{nd+1}$ and $\prod_{j=0}^n l_j(X) \mod X^{ nd+1}$.\\
As $H(X) = U(X) \cdot L(X) \mod X^{nd+1}$, we deduce that $d_H(X) = det(U(X)) \cdot det(L (X)) \mod X^{nd+1}$. Thus $H$ is also invertible and belongs to $\mathcal{R}^*$
We can note that as the $u_i(X)$ and $l_i(X)$ have invertible constant terms, the constant term of $d_H(X)$ is also invertible, thus $d_H(X)$ is invertible in $K[X]_{/X^{nd+1}}$.
\end{proof}

\begin{proposal}
$H(X)$ is random and uniform over $\mathcal{R}^*$.
\end{proposal}

\begin{proof}
We recall that $\mathcal{R} = \mathcal{M}_n(K\llbracket X \rrbracket_{X^{nd+1}})$
We define several sets.\\
Let $\mathcal{L} \subset \mathcal{R^*}$ be the set of invertible lower triangular matrices in $\mathcal{R}$.\\
Let $\mathcal{U} \subset \mathcal{R^*}$ be the set of invertible upper triangular matrices in $\mathcal{R}$.\\
Let $\mathcal{D} \subset \mathcal{R^*}$ be the set of invertible diagonal matrices in $\mathcal{R}$.
We will measure the size of our different sets.\\
For the set $K\llbracket X \rrbracket_{/X^{dn+1}}$, we have $nd+1$ coefficients which live in a field $K$ of size $q$. This set is therefore of size\\
$|K\llbracket X\rrbracket_{/X^{dn+1}}| = q^{nd+1}$\\

For the set $GL_n(K[X]_{/X^{dn+1}})$, we have the same set as the previous except the constant term which must be invertible. So we have $nd$ elements in $K$ and one element in $K^*$. The latter is of size $q-1$. So our set is size \\
$|GL_n(K[X]_{/X^{dn+1}})| = q^{nd}\cdot(q-1)$\\

For the set $\mathcal{R}$, we have a matrix of size $n\times n$ with coefficient in $K[X]_{/X^{dn+1}}$. So we have a set of size \\
$|\mathcal{A}| = (q^{nd+1})^{n^2} = q^{n^3d+n^2} $\\

For the set $\mathcal{L}$ and the set $\mathcal{U}$, we first notice that they are of identical sizes. Then we notice that they have $(\sum_{i=1}^n i ) - n = \frac{n(n-1)}{2}-n $ terms that live in $|K[X]_ {/X^{dn+1}}|$ plus the $n$ diagonal elements that live in $|GL_n(K[X]_{/X^{dn+1}})|$. So we have these two sets of size \\
$|\mathcal{L}| = |\mathcal{U}| = (q^{nd+1})^{\frac{n(n-1)}{2} } \cdot (q^{nd}\cdot (q-1))^n = (q^{\ frac{n^3d + n^2d + n^2 - n }{2}}) \cdot (q-1)^n$\\

Finally, the set $\mathcal{D}$ is composed of $n$ nonzero terms on its diagonal. So we have\\
$|\mathcal{D}| = q^{n^2d}(q-1)^n$\\

Let $h$ be the map $\mathcal{L}\times \mathcal{U} \longrightarrow \mathcal{R^*}$, $ L \times U \mapsto H$, and we denote $\mathcal{H} = h(\mathcal{L},\mathcal{U})$

\begin{lemma}
For each $H \in \mathcal{H} \quad|h^{-1}(H)| = |\mathcal{D}| $.
\end{lemma}
\begin{proof}
Let $H = LU$ and $D \in \mathcal{D}$. So, $LD^{-1} \in \mathcal{L}$ and $DU \in \mathcal{U}$. H therefore has at least $|\mathcal{D}|$ pre-images by $h$. \\
For each $H \in \mathcal{H} \quad|h^{-1}(H)| \geq |\mathcal{D}| $\\

Suppose there are two distinct $LU$ decomposition of a matrix $H$, such that $L_1 \neq L_0\cdot D^{-1}$ and $ U_1 \neq D\cdot U_0$.\\
$H = L_0\cdot U_0 = L_1\cdot U_1$, so we have that $L_1^{-1}L_0 = U_1U_0^{-1}$.\\
Knowing that the multiplication of two lower (resp. upper) triangular matrices returns a lower (resp. lower) triangular matrix, $L_1^{-1}L_0 \in \mathcal{L}$ and $U_1U_0^{-1} \in \mathcal{U}$.\\
Thus, knowing that $\mathcal{L} \cap \mathcal{U} = \mathcal{D}$, $L_1^{-1}L_0 = U_1U_0^{-1} = D \in \mathcal{D}$ . We can then write $L_1 = L_0\cdot D^{-1}$ and $ U_1 = D\cdot U_0$, which is false by hypothesis.\\
Thus, we have just proved that there does not exist a pre-image of $H$ which is not written like another pre-image of $H$ up to a diagonal matrix.\\
Thus, for each $H \in \mathcal{H} \quad|h^{-1}(H)| \geq |\mathcal{D}| $

\end{proof}

So we have $|\mathcal{H}| = \frac{|\mathcal{L}|\cdot|\mathcal{U}|}{|\mathcal{D}|}$
We can therefore count the proportion between the number of elements in $\mathcal{H}$ and the number of elements in $\mathcal{R^*}$

\begin{equation*}
\begin{split}
\frac{|\mathcal{H}|}{|\mathcal{R}|} & = \frac
     {|\mathcal{L}|\cdot|\mathcal{U}|}
     {|\mathcal{D}|\cdot |\mathcal{R}|}\\
 & = \frac
    {(q^\frac{n^3d + n^2d + n^2 - n }{2})^2 \cdot (q-1)^{2n}}
    {q^{n^2d}(q-1)^n \cdot q^{n^3d+n^2} }\\
 & = \frac
    {q^{n^3d + n^2d + n^2 - n } \cdot (q-1)^{2n}}
    {q^{n^2d} \cdot (q-1)^n \cdot q^{n^3d}\cdot q^{n^2} }\\
 & = \frac
    {q^{- n } \cdot (q-1)^{n}}{1}\\
 & = \left(\frac{q-1}{q}\right)^n\\
 &=\left(1 - \frac{1}{q}\right)^n
\end{split}
\end{equation*}
So if we can give a good lower for the comparison between $\mathcal{H}$ and $\mathcal{R^*}$:\begin{equation*}
\begin{split}
\frac{|\mathcal{H}|}{|\mathcal{R^*}|}
& \geq \left(1 - \frac{1}{q} \right)^n\\
& > 1 - \frac{n}{q} \\
\end{split}
\end{equation*}
We conclude that when $L$ and $U$ are chosen uniformly safe $\mathcal{L}\times \mathcal{U}$, then $H$ the output matrix of the protocol is distributed uniformly safe, $\mathcal{ H}$ which corresponds to almost all invertible matrices in the ring of series of truncated matrices of degree $nd+1$, when $n$ is negligible compared to $q$.

\end{proof}

\begin{center}
    \large\textbf{Complexity proofs}
\end{center}

\textbf{Round Complexity:}\\
Lines 2 and 3 of the algorithm require randomly and uniformly drawing $2n$ invertible polynomials of degree $dn$. This operation is carried out in constant number of rounds with the protocol of Bar-Ilan and Beaver \cite{BarIlan1989NoncryptographicFC}. As each draw is independent and carried out in parallel, the complexity is kept constant.\\
Line 5 uses the Fan-in Polynomial Multiplication protocol, the latter is proposed in constant round by Mohassel and Franklin \cite{efficient_Poly_operation}.\\
Lines 6 and 7 make use of the creation of a lower (respectively upper) triangular polynomial matrix split described earlier. It is therefore necessary to perform $O(n^2)$ draw uniform polynomials on $\mathcal{R}$. These draws are independent and can therefore be done in parallel, so the round complexity remains constant.\\
Line 8 uses a polynomial matrix multiplication protocol seen previously in constant number of rounds.\\
We conclude that the round complexity of this protocol is $O(1)$.\\

\textbf{Beaver triples on $K$}\\
Lines 2 and 3 require checking that the randomly drawn values are indeed invertible. For this, it is necessary to check that the constant coefficient of the drawn polynomials is indeed invertible. Bar-Ilan and Beaver \cite{BarIlan1989NoncryptographicFC} propose a protocol for this that requires a Beaver triple for verification. So we need $O(n)$ beaver triples for the $n$ iterations.\\
Line 5 proceeds to the execution of the Fan-In multiplication algorithm, which requires $O(dn^2)$ Beaver triples with the protocol of Mohassel and Franklin.\\
Lines 6 and 7 create the two desired matrices and do not require Beaver triples on $K$.\\
Line 8 corresponds to a polynomial matrix multiplication and therefore does not use beaver triples on $K$.\\
We therefore conclude that the number of Beaver triples on $K$ is $O(dn^2)$.\\

\textbf{Beaver triples on $\mathcal{R}$}\\
Line 8 requires the multiplication of a matrix polynomial. As seen previously, we need $O(1)$ Beaver triples of matrix polynomials.\\

\textbf{Complexity}\\
Lines 2 and 3 which want to generate $2n$ polynomials in $\mathcal{R^*}$, require for each invertibility check, an interactive multiplication. So we need to pass $O(n\cdot \log(q))$ bits per player.
Line 5, which uses the \textbf{Fan-In MultiplicationPoly} protocol proposed by Mohassel and Franklin, requires sending $O(dn^2\log(q))$bits of information.\\
Line 6 and 7 run internally and therefore do not require communication.\\
Line 8 which consists of a matrix multiplication of polynomials of degree at most $dn$ that is $O(dn^3\log(q))$. We therefore send $O(dn\cdot n^2 \log(q))$ bits.\\\textbf{Internal Complexity}\\
Lines 2 and 3 require $O(n)$ generations of invertible polynomials of degree $dn$. We therefore have a complexity in $O(dn^2+nN)$.
Line 5 which uses the Fan-In Multiplication Poly protocol of $n$ elements of degree $dn$ requires $O(n^3dN + n^2\mathsf{M}(dn)\log(dn)))$ .\\
Lines 6 and 7 require generating the two polynomial matrices, and thus to generate $O(n^2)$ polynomials of degree $dn$. So we have $O(dn^3)$ elements of $K$ to generate. So the internal complexity of this step is $O(dn^3)$.\\
Line 8 is to multiply two polynomial matrices. We therefore have a complexity in $O(\mathsf{MM}(n,nd)+n^3dN)$.\\
Knowing that $\omega < 3$ the internal complexity of this protocol is $O(\mathsf{MM}(n,nd)+n^3dN)$.\\

We can therefore conclude by summarizing here the complexities of this protocol.

$$
 \begin{array}{|c|c|c|c|c|}
 \hline
    \textbf{Round}&
    \textbf{B. Triples on $K$}&
    \textbf{B. Triples on $\mathcal{R}$}&
    \textbf{Communication} &
    \textbf{Internal}\\\hline
      O(1) & O(n^2d) & O(1) & O(n^3d\cdot \log(q)) & O(\mathsf{MM}(n,nd)+n^3dN)\\ \hline
 \end{array}
    $$
So we have a protocol to generate a mask. We will now see how to use it and compute the determinant.

\subsection{computation of the determinant of a matrix in $\mathcal{R^*}$}
We start from the scenario where the players have a secret sharing of a matrix $A(X) \in \mathcal{R^*}$, we then know that the constant coefficient of $A(X)$ is invertible. \\
The idea of this protocol is to mask the matrix $A$ thanks to the previous protocol, then to remove the mask once the determinant has been computed.\\
\begin{itemize}
    \item Players use the above protocol to generate $[H(X)]$ and $[d_H(X)]$.
    \item Players compute interactively, then reveal $[A(X)\cdot H(X)]$.
    \item Players publicly compute $e(X) = det(A(X)\cdot H(X))$
    \item $d_H(X)$ is an invertible polynomial in $K[X]_{/X^{nd+1}}$ by construction, so players compute the inverse with Newton's iteration protocol proposed by Mohassel and Franklin \cite{efficient_Poly_operation}.
    \item $[det(A)] = e / [d_H]$.
\end{itemize}

 \begin{algorithm}[H]
\SetAlgoLined
\SetKw{KwBy}{by}
\KwData{
$N$ players $P_1,\cdots,P_N$\\
$[A]$ the secret share of the matrix $A \in \mathcal{R^*}$
}
\KwResult{
$[d_A(X)] \in K[X]_{/X^{nd+1}}$ the split of the determinant of $A(X)$}
\vspace{5pt}

$P_i: ([H]_i,[d_H(X)]_i) \gets \textbf{RandMatPolyDet}(n,d)$\\
$P_i: [d_H^{-1}(X)]_i \gets $\textbf{InversePoly}$( P_i: [d_H(X)]_i)$\\
$P_i: [E(X)]_i \gets $\textbf{MultiplicationMatricesPoly}$( P_i: [A(X)]_i,[H(X)]_i)$\\
\textbf{Reveal}$( P_i : [E(X)]_i)$\\
$e(X) \gets $\textit{Determiner}$(E(X))$\\
$P_i:[d_A(X)]_i \gets P_i:e(X)\cdot [d_H^{-1}(X)]_i$\\

\textbf{Return }$P_i: [d_A(X)]_i$
\caption{\textbf{DetInv}$((P_i : [A(X)]_i)_{i\in\llbracket 1,N\rrbracket})$}
\end{algorithm}

\begin{center}
    \large\textbf{Proof of correction}
\end{center}
\begin{proposal}
The output of the \textbf{DetInv} protocol is a split of the determinant of $A(X)$, $[det(A(X))]$
\end{proposal}
\begin{proof}
The protocol computes the determinant of the polynomial matrix $E(X)$ with an algorithm for computing the determinant of a polynomial matrix.\\
We have $E(X) = A(X)\cdot H(X)$. Thus $e(X) = det(E(X)) = det(A(X)) \cdot det(H(X)) = d_A \cdot d_H$. We also know that the determinant of $A(X)$ and the determinant of $H(X)$ are invertible in $K[X]_{/X^{dn+1}}$. Thus, we can compute the partition of the inverse of $d_H$ with the algorithm of inversion by the iteration of newton proposed by Mohassel and Franklin.\\
Finally, players internally compute $[d_A] = e \cdot [d_H^{-1}]$.
We have the correction knowing that $deg(det(A(X))) \leq nd $ and
$$A(X) = \sum_{i = 0}^{N}[d_A(X)]_i = \sum_{i = 0}^{N} e(X) \cdot [d_H^{-1} (X)] = e(X)\sum_{i = 0}^{N} [d_H^{-1}(X)] = e(X)\cdot d_H^{-1}(X) $$
\end{proof}
\begin{proposal}
The \textbf{DetInv} protocol does not disclose any information about $A(X)$
\end{proposal}
\begin{proof}
We know that $A(X),H(X) \in \mathcal{R^*}$, and that $H(X)$ is drawn uniformly and randomly in $\mathcal{H}$ such that $\mathcal {H} \subset \mathcal{R^*}$ and $\frac{\mathcal{H}}{\mathcal{R^*}}\geq 1 - n/q$. Thus, if $n$ is negligible compared to $q$, we can consider that $H(X)$ is drawn uniformly on $\mathcal{R^*}$. Thus, $E(X)$ is equivalent to a matrix drawn randomly and uniformly by a random variable on $\mathcal{R^*}$ (Lemma 1). We cannot therefore derive any information from the knowledge of $E(X)$. The rest of the protocol consists of known protocols that do not disclose information about $A(X)$.
\end{proof}

\begin{center}
    \large\textbf{Complexity Proof}
\end{center}

\textbf{Round Complexity:}\\
The first line uses the \textbf{RandMatPolyDet} protocol which is in constant round. \\
The second line uses the inversion protocol with known newton iteration in constant round \cite{efficient_Poly_operation}.\\
The third line asks for an interactive polynomial matrix multiplication protocol, we have a constant round protocol for that.\\
The fourth line proposes to reveal $E(X)$, each participant sends his secret share in the communication channel and the number of rounds remains constant. \\
The other lines do not use protocols that involve communications.
So we have a protocol in constant round.\\
\textbf{Beaver triples on $K$:}\\
The first line calls for \textbf{RandMatPolyDet}, so we need $O(n^2d)$ Beaver triples on $K$.\\
Line 2 corresponds to Newton's iteration which requires $O(dn)$ Beaver triples.\\
Line 3 is a polynomial multiplication of two matrices and therefore requires no Beaver triples on $K$.\\
The \textbf{Reveal} line 4 does not require Beaver triples on $K$.\\
The other lines do not use interactive protocols and therefore do not require any Beaver triples.
So we need a total of $O(n^2d)$ Beaver triples for the protocol.\\

\textbf{Beaver triples on $\mathcal{R}$:}\\
\textbf{RandMatPolyDet} therefore requires $O(1)$ of Beaver triples on $\mathcal{R}$.\\
The protocol from newton's iteration to the second line does not require any matrix multiplication, so the protocol does not require Beaver triples on $\mathcal{R} $.\\
Line 3 is matrix polynomial multiplication and therefore requires $O(1)$ Beaver triples over $\mathcal{R} $.\\
The \textbf{Reveal} line 4 does not require Beaver triples on $K$.\\
the other lines are not interactive, so we have a protocol that requires $O(1)$ Beaver triples over $\mathcal{R}$.\\

\textbf{Communication complexity:}\\
\textbf{RandMatPolyDet} requires $O(n^3d\log(q))$communication bits.\\
The second line refers to a matrix polynomial multiplication and therefore requires $O(n^3d\log(q))$ communication bits\\
Revealing the $E(X)$ matrix in the third row requires each player to send a polynomial matrix of degree at most $dn+1$ into the communication channel. Thus, the communication complexity is $O(n^3d\log(q))$bits.\\
The $5^{eme}$ line uses Newton's iteration which requires $O(dn\log(q))$communication bits.\\
So we have a protocol that commits $O(dn^3\log(q))$ communication bits per player.\\

\textbf{Internal complexity:}\\
\textbf{RandMatPolyDet(n,d)} has an internal complexity of $O(\mathsf{MM}(n,nd)+n^3dN)$.\\
The second line uses the \textbf{InversePoly} protocol proposed by Mohassel and Franklin in complexity $O(ndN + M(nd)\log(nd))$ to invert a polynomial of size $nd$.\\
The third line consists of multiplying two polynomial matrices of degree $dn$. The protocol requires a complexity of $O(\mathsf{MM}(n,nd)+n^3dN)$.
Line 4 is just to send a data in the communication channel and to reconstitute the polynomial matrix, so $O(n^3dN)$ additions per player.\\
The computation of the determinant of a polynomial matrix can be reduced to the computation of the matrix polynomial product \cite{storjohann} and therefore has a complexity of $O(\mathsf{MM}(n,nd))$
So we finally have an internal complexity in $O(\mathsf{MM}(n,nd)+n^3dN)$

$$
 \begin{array}{|c|c|c|c|c|}
 \hline
    \textbf{Round}&
    \textbf{B. Triples on $K$}&
    \textbf{B. Triples on $\mathcal{R}$}&
    \textbf{Communication} &
    \textbf{Internal}\\\hline
     O(1) & O(n^2d) & O(1) & O(n^3d\cdot \log(q)) & O(\mathsf{MM}(n,nd)+n^3dN)\\ \hline
 \end{array}
$$

This protocol makes it possible to compute the determinant of a matrix when the latter is known to be invertible. Nevertheless, if we use this protocol on a matrix $A(X)$ which is not necessarily invertible in the truncated series. Then the information of the singularity of $A(X)$ will be immediately revealed. Indeed, when computing $e(X)$ if its constant term is equal to 0, we know that $det(A(0)\cdot H(0)) = det(A(0)) \cdot det (H(0)) = 0 $, or $det(H(0)) \neq 0$ by definition of $H$. So, we learn that $det(A(0)) = 0$ is that therefore $A(X)$ is not invertible in truncated series. We therefore reveal information, which is problematic.\\

To overcome this problem, which is found in conventional matrices, Cramer and Damgard propose a second protocol.
\subsection{computation of the determinant of a matrix in $\mathcal{R}$ }

We now assume that $A \in \mathcal{R} (\mathcal{R} = \mathcal{M}_n(K\llbracket X \rrbracket))$
    
We know that the determinant of any matrix is determined by the constant coefficient of its characteristic polynomial. For a matrix $M$ of dimension $n,n$, $f_M(0) = (-1)^n det(M)$. The idea here is to compute the evaluation of the characteristic polynomial in $n+1$ points then to find the determinant in the constant term.

    \begin{itemize}
        \item Players publicly randomly and uniformly draw $n+1$ distinct polynomials of degree at most $nd$. We denote $z_i(X)$ on $K[X]_{/X^{nd+1}}$, $\forall i,j \quad z_i(X)-z_j(X) \neq 0$. The data is immediately revealed.\\
        If some $z_i(X)$ are not distinct, we can relaunch the $z_i(X)$ which cause the problem.
        \item Players internally compute $[z_i(X)I_n - A(X)]$ for $0 \leq i \leq n$.
        \item Players get $[det(z_i(X)I_n - A(X))]$ for $0 \leq i \leq n$ thanks to the protocol for computing the determinant of a matrix in $\mathcal{A^* }$ seen previously.
        \item Players then compute
        $$l_i = \prod^{n}_{j\neq i, j=0} \frac{0 - z_j}{z_i - z_j}$$
        $$[det(A(X))] = (-1)^n \sum^{n}_{i=0} l_i [det(z_i(X)I_n - A(X))]$$
    \end{itemize}
    
    We know that $z_i(X)I_n - A(X)$ is invertible if and only if $z_i(X)$ is not an eigenvalue of $A(X)$.

 \begin{algorithm}[H]
\SetAlgoLined
\SetKw{KwBy}{by}
\KwData{
$N$ players $P_1,\cdots,P_N$\\
$[A(X)]$ a shared polynomial matrix\\
}
\KwResult{$[D(X)] \in K[X]$ the split of the determinant of $A(X)$}
\vspace{5pt}

\ForEach{$j \in \llbracket 0,n \rrbracket$ }
{
$P_i : z_j(X) \gets \textit{RandomPoly}(nd)$\textit{ such as $\forall j,k \quad z_k - z_j \neq 0$}\\
$z_j(X) \gets \textbf{Reveal}(P_i = [z_j(X)]_i)$\\

}
\ForEach{$j \in \llbracket 0,n \rrbracket$ }
{
$P_i: [z_j(X)I_n - A(X)]_i \gets P_i:z_j(X)\cdot I_n - [A(X)]_i$\\
$P_i: [d_j] \gets \textbf{DetInv}(P_i: [z_j(X)I_n - A(X)]_i)$\\
$l_j(X) \gets \prod^{n}_{k\neq j, k=0} \frac{0 - z_k(X)}{z_j(X) - z_k(X)}$\\
}

$P_i : [D(X)]_i \gets (-1)^n \sum_{k=0}^n [d_k]_i\cdot l_k(X) $\\
\textbf{Return:}$P_i: [D(X)]_i$

\caption{DetMatPoly$((P_i : [A(X)]_i)_{i\in\llbracket 1,N\rrbracket})$}
\end{algorithm}

\begin{center}
    \textbf{Complexity proof}\\
\end{center}

\textbf{Round complexity: }\\
Each of the interactive protocols are known in constant round. They are all in a loop, but since all the executions are independent of each other, the protocols can be parallelized and therefore be in constant round. \textbf{DetMatPoly} is therefore in constant round.\\

\textbf{Beaver triples on $K$: }\\
In the first loop, the interactive protocols do not consume Beaver triples. In the second loop, the $\textbf{DetInv}$ protocol consumes $O(n^2d)$ Beaver triples out of $K$ per execution, or $O(n^3d)$ triples in total.\\

\textbf{Beaver triples on $\mathcal{R}$: }
As before, the only protocol that consumes Beaver triples is \textbf{DetInv} which consumes $O(1)$ triplet over $\mathcal{R}$ per execution, for a total of $O(n)$.\\
\textbf{Communication complexity: }\\
The \textbf{Reveal} protocol of $n$ polynomials of degree at most $dn$ requires a communication of $O(n^2d \log(q))$ bits per player for the $n$ executions. The \textbf{DetInv} protocol requires $O(n^3d\log(q))$ bits per execution, totaling $O(n^4\log(q))$ communication bits.\\

\textbf{Internal complexity: }\\
It is necessary to generate $n$ polynomials of degree $nd$ which gives $O(n^2d)$ operations. The $n$ operations of \textbf{Reveal} require $O(Nn^2d)$ additions. Knowing that $n$ is small compared to $|K| = q$, we can neglect the few $z_i$ that we raise to have them all distinct.\\
The protocol \textbf{DetInv} at line 7, asks for its $n$ executions $O(n\cdot \mathsf{MM}(n,nd)+n^4dN)$. Finally, the interpolation on line 8 and 10 requires $O(n^2\mathsf{M}(n))$ operations knowing that the evaluation points are polynomials. So we have a final complexity in $O(n\cdot \mathsf{MM}(n,nd)+n^4dN)$

$$
 \begin{array}{|c|c|c|c|c|}
 \hline
    \textbf{Round}&
    \textbf{B. Triples on $K$}&
    \textbf{B. Triples on $\mathcal{R}$}&
    \textbf{Communication} &
    \textbf{Internal}\\\hline
      O(1) & O(n^3d) & O(n) & O(n^4d\cdot log(q)) & O(n\cdot \mathsf{MM}(n,nd)+n^4dN) \\hline
     
 \end{array}
$$

\begin{proposal}
Algorithm 7 does return the determinant of $A$
\end{proposal}
\begin{proof}
Let $f_A(X,Y)$ be the characteristic polynomial of $A(X)$. We know that $det(A(X)) = (-1)^n \cdot f_A(X,0)$. We therefore evaluate the characteristic polynomial at $f_A(X,z_i(X))$ with steps 6 and 7. Finally, we interpolate the polynomial then evaluate it at $f_A(X,0)$. We therefore return a partition of the determinant of $A(X)$.
$$l_j(X) = \prod^{n}_{k\neq j, k=0} \frac{0 - z_k(X)}{z_j(X) - z_k(X)} $$
$$[det(A(X))] = (-1)^n \cdot [f_A(X,0)] = (-1)^n \cdot \sum_{k=0}^n [d_k]_i \cdot l_k(X) $$
\end{proof}

As far as security is concerned, the protocol does not reveal any information about $A(X)$ knowing that the only protocol which engages the security of $A(X)$ during communication is \textbf{DetInv}. This last protocol compromises the security of $A(X)$ if the matrix $z_jI_n - A(X)$ is not invertible. In the latter case then, the \textbf{DetInv} protocol will reveal that $z_j$ is an eigenvalue of the characteristic polynomial of $A(X)$. Only if $q$ is large in front of $n$ then the probability is low.\\

So we have a second protocol for computing the determinant of a polynomial matrix.\\

\section{Method$\mod f$ an irreducible polynomial}

A last method considered computing our determinant would be to transform the ring of polynomial matrices of degree at most $d$ into the same ring modulo an irreducible polynomial $f$ of degree greater than $dn$. We will then find ourselves in a ring of coefficient matrix in a finite field, and then the protocols of Cramer and Damgard provided for this purpose become perfectly functional.\\

We will not describe these last protocols, because we have just translated the direct adaptation of them in the last section, the idea is the same for matrices with coefficients in a finite field. We will only focus on the complexity of this method and therefore its potential interest.\\

First, let's talk about the irreducible polynomial $f$.
There are several computation methods to obtain such a polynomial, the best known of which are probabilistic. These methods consist in testing the irreducibility of a polynomial drawn randomly (for example thanks to the Berlekamp algorithm), and this, until falling on an irreducible polynomial.
It is therefore a Las Vegas-type algorithm. \\
One can also be satisfied with the use of families of irreducible polynomials and agree to take a slightly greater degree, even if it means having a little greater arithmetic complexity afterwards.
We will simply note $\mathsf{H}(n)$ the complexity for the computation of an irreducible polynomial of degree $n$.\\

As discussed earlier, we will not detail protocols, only complexities. We need to define the cost of arithmetic operations in the field $K[X]_{/f}$ in terms of the cost in the field $K$. Let $dn+1$ be the degree of $f$, knowing that the determinant has degree at most $dn$, addition costs $O(dn)$ , multiplication costs $O(\mathsf{M(dn )})$. \\
\begin{center}
    \textbf{Cramer and Damgard protocol complexity in $\mathcal{M}_n(K[X]_{/f})$, determining case $D(0) \neq 0$: }
\end{center}

\textbf{Round complexity:}\\
The computation of the irreducible polynomial $f$ is done in public.\\
The initial protocol is in constant round. The arithmetic operations in the field $K[X]_{/f}$ are done without interaction for the addition and the multiplication by a public data, and in constant round for the multiplication of secrets. Thus, the transformed protocol is in constant round.\\
\textbf{Beaver triples on $K[X]_{/f}$:}\\
The computation of the mask in the Cramer and Damgard protocol costs $O(n)$ Beaver triples over $K$. Its adaptation then requires $O(n)$ triples on $K[X]_{/f}$. The rest of the protocol does not use any.\\
\textbf{Beaver triples on $\mathcal{M}_n(K[X]_{/f})$:}\\
Cramer and Damgard's protocol uses two matrix Beaver triples in its protocol. Its adaptation will therefore require $O(1)$ triples in $\mathcal{M}_n(K[X]_{/f})$.\\
\textbf{Communication complexity:}\\
Cramer and Damgard's protocol requires $O(n^2\log(q))$ communication bits per player because of matrix multiplication, which involves revealing a matrix. Thus, by multiplying polynomial matrices we will need $O(n^3d\log(q))$ communication bits per player.\\
\textbf{Internal complexity:}\\
Computing the irreducible polynomial costs $\mathsf{H}(nd)$
The Cramer and Damgard protocol requires $O(n^2N + \mathsf{MM}(n))$ arithmetic operations. $O(n^2N)$ for communications reconstructions, and $\mathsf{MM}(n)$ for interactive multiplications. Its adaptation therefore requires $O(n^3dN + \mathsf{MM}(n,nd)+\mathsf{H}(nd))$.
 
\begin{center}
    \textbf{Complexity of the Cramer and Damgard protocol in $\mathcal{M}_n(K[X]_{/f})$, general case of the determinant: }
\end{center}

\textbf{Round complexity:}\\
The protocol for computing the determinant in the case different from zero is executed $n$ times in parallel, the rest of the operations are known in constant round or without interaction. The protocol is therefore in constant round.\\
\textbf{Beaver triples on $K[X]_{/f}$:}\\
The last protocol used $O(n)$ Beaver triples on $K[X]_{/f}$, we repeat this protocol $n+1$ times and the rest of the operations do not require Beaver triples. So we need $O(n^2)$ Beaver triples on $K[X]_{/f}$.\\
\textbf{Beaver triples on $\mathcal{M}_n(K[X]_{/f})$:}\\
The last protocol used $O(1)$ Beaver triples on $\mathcal{M}_n(K[X]_{/f})$, we repeat this protocol $n+1$ times and the rest of the operations do not require Beaver triples. We therefore need $O(n)$ Beaver triples on $\mathcal{M}_n(K[X]_{/f})$.\\
\textbf{Communication complexity:}\\
The last protocol used $O(n^3d\log(q))$ communication bits, we repeat this protocol $n+1$ times and the \textbf{Share}() of $n$ polynomials of degree at most $ nd$ requests $O(n^2d\log(q))$ bits. We therefore need $O(n^4d\log(q))$ communication bits.\\
\textbf{Internal complexity:}\\
The last protocol used $O(n^3dN + \mathsf{MM}(n,nd)+\mathsf{H}(nd))$ arithmetic operations in $K$, we repeat this protocol $n+1$ times. We therefore need $O(n^4dN + n\cdot\mathsf{MM}(n,nd)+\mathsf{H}(nd))$. The cost of the operation of the interpolation in public point of a polynomial with coefficient in $K[X]_{/f}$, evaluated in points which are polynomials, costs as in the previously proposed protocol $O (n^2\mathsf{M(nd)})$.\\

\section{Comparison of different protocols}
We therefore have different protocols to compute the determinant of a polynomial matrix, we compare depending on the case where the constant coefficient of the determinant is known, invertible or not
{\footnotesize
$$
 \begin{array}{|c|c|c|c|}
 \hline

  \multicolumn{4}{|c|}{\cellcolor{Silver}\textbf{Determinant of a polynomial matrix, case $det(A(0)) \neq 0$}} \\\hline
  
   \textbf{Method} & \textbf{Eval/interpol} & \textbf{modulo $X^{nd+1}$} & \textbf{modulo $f$} \\\hline
    \textbf{Round}& O(1) & O(1) &O (1) \\\hline
    \textbf{B. Triples on $K$}& O(n^2d) & O(n^2d) & - \\\hline
    \textbf{B. Triples on $K[X]_{<nd}$}& - & - & O(n) \\\hline
    \textbf{B. Triples on $\mathcal{M}_n(K)$}& O(nd) & - & -\\\hline
    \textbf{B. Triples on $\mathcal{M}_n(K[X]_{<nd})$}& - & O(1) & O(1)\\\hline
    \textbf{Communication} & O(n^3d\cdot \log(q)) & O(n^3d\cdot \log(q))& O(n^3d\cdot \log(q)) \\\hline
    \textbf{Internal}
    & O(n^3dN + nd\cdot \mathsf{MM}(n)+ n^3d^2 )
    & O(n^3dN + \mathsf{MM}(n,nd))
    & O(n^3dN + \mathsf{MM}(n,nd)+\mathsf{H}(nd))\\\hline
    \hline
    
     \multicolumn{4}{|c|}{\cellcolor{Silver}\textbf{Determinant of a polynomial matrix, general determinant case}} \\\hline
  
   \textbf{Method} & \textbf{Eval/interpol} & \textbf{modulo $X^{nd+1}$} & \textbf{modulo $f$} \\\hline
    \textbf{Round}& O(1) & O(1) &O (1) \\\hline
    \textbf{B. Triples on $K$}& O(n^3d) & O(n^2d) & - \\\hline
    \textbf{B. Triples on $K[X]_{/f}$}& - & - & O(n^2) \\\hline
    \textbf{B. Triples on $\mathcal{M}_n(K)$}& O(n^2d) & - & -\\\hline
    \textbf{B. Triples on $\mathcal{M}_n(K[X]_{/f})$}& - & O(n) & O(1)\\\hline
    \textbf{Communication} & O(n^4d\cdot \log(q)) & O(n^4d\cdot \log(q))& O(n^4d\cdot \log(q)) \\\hline
    \textbf{Internal}
    & O(n^4dN + n^2d\cdot \mathsf{MM}(n)+ n^3d^2 )
    & O(n^4dN + n\mathsf{MM}(n,nd))
    & O(n^4dN + n\cdot\mathsf{MM}(n,nd)+\mathsf{H}(nd))\\ \hline
    \end{array}
$$
  }
  
  We thus have three protocols for computing the determinant.
  All three are relatively similar in complexity.\\
  
  We can say that even if the method modulo $f$ and modulo $X^{nd+1}$ do not use the same types of Beaver triples, they use in total a comparable volume $O(n^4d )$. The triplets being managed in a pre-computation phase, we can conclude that the difference is made on the computation of the irreducible polynomial which is probabilistic \cite{ModernCA}.\\ We draw a polynomial of degree, $nd$ then we check if it is irreducible in $O(\mathsf{M}(n)\log(q) + (n^{(\omega+1)/2} + n^{1/2}\mathsf{M} (n))\delta(n)\log(n))$ with $\delta(n)$ the number of prime factors of $n$. We expect to find an irreducible polynomial on average in $O^ (n^{(\omega+3)/2} + n^2\log(q))$.
  
  We can compare the protocol modulo $X^{nd+1}$ and the method by interpolation evaluation.\\
  We take the Bostan-Schost algorithm \cite{Bostan2010} for polynomial matrix multiplication and the fast Fourier transform for polynomial multiplication.
  
  We then have $O(n^4dN + n\mathsf{MM}(n,nd)) = O(n^4dN + \mathsf{MM}(n)\cdot nd + n^2\mathsf{M}( nd)) = O(n^4dN + \mathsf{MM}(n)\cdot nd + n^3d\cdot \log(nd))$ for the computation of the determinant with the modulo method $X^{nd+1 }$.
  
  The interpolation evaluation protocol has a complexity of ($O(n^4dN + \mathsf{MM}(n)\cdot n^2d+ n^3d^2 )$). So whether $n$ is very large in front of $d$ or vice versa, the protocol with method $X^{nd+1}$ has an asymptotically more efficient complexity.\\
  
 This difference comes from the use of polynomial matrix Beaver triples. The method of triples actually hides part of the computations in the form of pre-computation upstream of the protocols.
  It is therefore not completely surprising to see the evaluation protocol a little less efficient.\\
  
  We therefore conclude that the modulo method $X^{nd+1}$ has the most interesting arithmetic complexity.
  
  However, this result needs to be qualified. Indeed, the proposed protocols from Cramer and Damgard\cite{Cramer2001SecureDL} have some problems.
  
  The protocol for computing the determinant in the general case, for example, can reveal eigenvalues of the matrix $A(X)$. If the chosen value $z_i$ is an eigenvalue of $A(X)$, then the \textbf{DetInv} protocol will reveal that $det(z_i(X)I_n - A(X)) = 0$. However, we have not had time to address this aspect in this report.\\
  
  The second problem which arises is the quasi-uniformity of the mask $H(X)$ used in the first protocol for computing the determinant. We could see that the matrix $H$ is not uniform in $\mathcal{R^*}$, but in $\mathcal{H}$.\\
  
   Thus, by drawing a random value in $\mathcal{H}$ we have a probability $\frac{1}{|\mathcal{H}|} > \frac{1}{|\mathcal{R^*}| } $ to find the right mask and thereby find $A(X)$. Information about $A(X)$ has thus been leaked, even if when $q >> n $ this information is negligible.\\
  
   The bound found on $\frac{|\mathcal{H}|}{|R|}$ in this report in the framework of the method modulo $X^{nd+1}$ is exactly the same as in the framework of a matrix with coefficients in a finite field. Therefore, the methods cannot be compared. The terminal is a little too wide to observe the differences.

%% file: 3-Conclusion.tex
\chapter{Conclusion}
During this internship, we studied different existing secure multiparty computation protocols. They allowed us to manipulate elements in a finite field, matrices, or even polynomials.

We were thus able to propose protocols for manipulating shared polynomial matrices.
We have, by adapting the method used by Cramer and Damgard, proposed three methods for computing the determinant of a shared polynomial matrix. We have proposed a method which evaluates, then interpolates the determinant of the polynomial matrix and then two methods to modify the ring of matrices with polynomial coefficients $\mathcal{M}_n(K[X])$, to be able to adapt the ideas of Cramer and Damgard.\\
We then compared them, in several forms of complexity and on the security of the shared data.\\
It has been observed that the different protocols are ultimately similar on the different levels of complexity and security. \\
Nevertheless, the modulo $X^{nd+1}$ method stands out from the others and happens to be the most efficient of the three methods studied during this internship.\\

The objective of this internship was to be able to define, then compare, several adaptations of the determinant computation protocol, for a coefficient matrix in a finite field. The arithmetic complexity, required by the protocols of multipart computation, is an aspect which is little approached in the various articles studied. Round complexity is generally preferred, as it is considered more time-consuming than arithmetic complexity. Keeping a constant complexity in rounds was therefore an objective. Nevertheless, the arithmetic complexity is often neglected in classical protocols, which is perhaps an impediment to implementation, so we laid out those complexities.\\
 
In the context of coefficient matrices in a finite field, the computation of the determinant could be improved, in particular by the computation of the characteristic polynomial with the Leverrier method, which we were not able to approach during this internship.\\
We could then seek to adapt these methods to the case of polynomial matrices in the future.

%% file: Complexite_annexe.tex
\begin{landscape}
{
\section{Complexities of the protocols studied}
\tiny
$$
\begin{array}{|c|c|c|c|c|c|}
\hline
\multicolumn{5}{|c|}{\cellcolor{Silver}\textbf{Generation of Beaver triples}} \\\hline
    \textbf{Protocol} &
    \textbf{Round}&
    \textbf{B. Triples on $K$}&
    \textbf{Communication} &
    \textbf{Internal} \\\hline

    \textbf{$BT_K$}& \multicolumn{4}{|c|}{\textit{Oracle}}\\\hline
    \textbf{$BT_{K[X]_{<d}}$}& O(1) & O(d) & O(d\log(q)) & O(dN+d\mathsf{M}( d))\\\hline
    \textbf{$BT_{\mathcal{M}_n(K)}$}& O(1) & O(n^\omega) & O(n^\omega\log(q)) & O(n^ \omega N)\\\hline
    \textbf{$BT_{\mathcal{M}_n(K[X]_{<d})}$}& O(1) & O(dn^\omega) & O(dn^\omega \log(q )) & O(dn^\omega N+d^2n^2)\\\hline

    \end{array}
    $$$$
    \begin{array}{|c|c|c|c|c|c|}
    \hline
    
    \multicolumn{5}{|c|}{\cellcolor{Silver}\textbf{Protocols on secretly shared $K$ elements}} \\\hline
     \textbf{Protocol} &
    \textbf{Round}&
    \textbf{B. Triples on $K$}&
    \textbf{Communication} &
    \textbf{Internal} \\\hline
    
    \textbf{Reveal}& O(1) & - & O(\log(q)) & O(N)\\\hline
    \textbf{Multiplication}&
    O(1) & O(1) & O(\log(q))& O(N)\\\hline
    \textbf{RandInv}&
    O(1) & O(1) & O(\log(q)) & O(N)\\\hline
    \textbf{Reverse} &
    O(1) & O(1) & O(\log(q)) &O(N) \\\hline
    \textbf{Fan-In Multiplication ($t$ items)} &
    O(1) & O(t) & O(t\log(q)) &O(Nt) \\\hline

        \end{array}
    $$$$
    \begin{array}{|c|c|c|c|c|c|}
    \hline
        \multicolumn{6}{|c|}{\cellcolor{Silver}\textbf{Protocols on elements of $\mathcal{M}_n(K)$ shared secretly}} \\\hline
     \textbf{Protocol} &
    \textbf{Round}&
    \textbf{B. Triples on $K$}&
    \textbf{B. Triples on $\mathcal{M}_n(K)$}&
    \textbf{Communication} &
    \textbf{Internal} \\\hline
    \textbf{Reveal}&
    O(1) & - & - & O(n^2\log(q)) &O(n^2N)\\\hline
    \textbf{MulMat}&
    O(1) & - & O(1) & O(n^2\log(q)) & O(n^2N+n^{\omega})\\\hline
    \textbf{RandMatInv}&
    O(1) & - & O(1) & O(n^2\log(q)) & O(n^2N + n^{\omega})\\\hline
    \textbf{InverseMat} &
    O(1) & - & O(1) & O(n^2\log(q)) & O(n^2N + n^{\omega})\\\hline
    \textbf{Fan-In Multiplication ($t$ items)} &
    O(1) & - & O(t) & O(t\cdot n^2\log(q)) & O(tn^2N + tn^{\omega})\\\hline
    \textbf{RandDetMat} &
    O(1) & O(n) & O(1) & O(n^2\log(q)) & O(n^2N + n^{\omega})\\\hline
    \textbf{Computation of the determinant (invertible case)} &
    O(1) & O(n) & O(1) & O(n^2\log(q)) & O(n^2N + n^{\omega})\\\hline
    \textbf{Computation of the determinant (general case)} &
    O(1) & O(n^2) & O(n) & O(n^3\log(q)) & O(n^3N + n^{\omega+1})\\\hline

    \end{array}
    $$$$
    \begin{array}{|c|c|c|c|c|}
    \hline
      \multicolumn{5}{|c|}{\cellcolor{Silver}\textbf{Protocols on elements of $K[X]$ of degree $n$ shared secretly}} \\\hline
    
     \textbf{Protocol} &
    \textbf{Round}&
    \textbf{B. Triples on $K$}&
    \textbf{Communication} &
    \textbf{Internal} \\\hline

     \textbf{Rating}&
     - & - & - & O(n)\\\hline
    \textbf{Interpolation in known points} &
     - & - & - & O(\mathsf{M}(n)\log(n)) \\\hline
    \textbf{Reveal}&
     - & - & - & O(nN)\\\hline
    \textbf{MulPoly}&
     O(1) & O(n) & O(n\log(q)) & O(nN + \mathsf{M}(n)\log(n))\\\hline
    \textbf{InversePoly} &
    O(1) & O(n) & O(n\log(q)) & O(nN + \mathsf{M}(n)\log(n))\\\hline
    \textbf{Fan-In Multiplication ($t$ items)} &
    O(1) & O(t^2 n) & O(t^2n\log(q)) & O(t^2nN+ t^2\mathsf{M}(n)\log(n) + \mathsf{ H}(n))\\\hline
    \textbf{EuclideanDiv} &
    O(1) & O(n) & O(n\log(q)) & O(nN + \mathsf{M}(n)\log(n))\\\hline
    \textbf{Interpolation in secret points} &
    O(1) & O(n^2) & O(n^2\log(q)) & O(n^2N+ n\mathsf{M}(n)\log(n) + \mathsf{H}( n))\\\hline
    \end{array}
    $$$$
    \begin{array}{|c|c|c|c|c|c|c|}
    \hline
      \multicolumn{7}{|c|}{\cellcolor{Silver}\textbf{Protocols on elements of $\mathcal{M}_n(K[X])$ of degree $d$ shared secretly}}\\\ hline
    \textbf{Protocol} &
    \textbf{Round}&
    \textbf{B. Triples on $K$}&
    \textbf{B. Triples on $\mathcal{M}_n(K)$}&
    \textbf{B. Triples on $\mathcal{M}_n(K[X]_{<d})$}&
    \textbf{Communication} &
    \textbf{Internal} \\\hline

     \textbf{Rating}&
     - & - & -&- & - & O(n^2d)\\\hline
    \textbf{Interpolation in known points}&
     - & - & -&- & - & O(n^2d) \\\hline
    \textbf{Reveal}&
     O(1) & - & - &-& O(n^2d\log(q)) & O(n^2dN)\\\hline
    \textbf{MulPolyMat}&
     O(1) & - &-& O(1) & O(n^2d\log(q)) & O(n^2dN + \mathsf{MM}(n,d))\\\hline
     
     \textbf{RandPolyMatInv}_{/X^d} &
     O(1) & - & O(1) & - & O(n^2\log(q)) & O(n^2N + \mathsf{MM}(n)) \\\hline
      
\end{array}$$
}
\end{landscape}

{\footnotesize

$$
 \begin{array}{|c|c|c|c|}
 \hline

  \multicolumn{4}{|c|}{\cellcolor{Silver}\textbf{Determinant of a polynomial matrix, determining case $\neq 0$}} \\\hline
  
   \textbf{Method} & \textbf{Eval/interpol} & \textbf{modulo $X^{nd+1}$} & \textbf{modulo $f$} \\\hline
    \textbf{Round}& O(1) & O(1) &O (1) \\\hline
    \textbf{B. Triples on $K$}& O(n^2d) & O(n^2d) & - \\\hline
    \textbf{B. Triples on $K[X]_{<nd}$}& - & - & O(n) \\\hline
    \textbf{B. Triples on $\mathcal{M}_n(K)$}& O(nd) & - & -\\\hline
    \textbf{B. Triples on $\mathcal{M}_n(K[X]_{<nd})$}& - & O(1) & O(1)\\\hline
    \textbf{Communication} & O(n^3d\cdot \log(q)) & O(n^3d\cdot \log(q))& O(n^3d\cdot \log(q)) \\\hline
    \textbf{Internal}
    & O(n^3dN + nd\cdot \mathsf{MM}(n)+ n^3d^2 )
    & O(n^3dN + \mathsf{MM}(n,nd))
    & O(n^3dN + \mathsf{MM}(n,nd)+\mathsf{H}(nd))\\\hline
    \hline
    
     \multicolumn{4}{|c|}{\cellcolor{Silver}\textbf{Determinant of a polynomial matrix, general determinant case}} \\\hline
  
   \textbf{Method} & \textbf{Eval/interpol} & \textbf{modulo $X^{nd+1}$} & \textbf{modulo $f$} \\\hline
    \textbf{Round}& O(1) & O(1) &O (1) \\\hline
    \textbf{B. Triples on $K$}& O(n^3d) & O(n^2d) & - \\\hline
    \textbf{B. Triples on $K[X]_{/f}$}& - & - & O(n^2) \\\hline
    \textbf{B. Triples on $\mathcal{M}_n(K)$}& O(n^2d) & - & -\\\hline
    \textbf{B. Triples on $\mathcal{M}_n(K[X]_{/f})$}& - & O(n) & O(1)\\\hline
    \textbf{Communication} & O(n^4d\cdot \log(q)) & O(n^4d\cdot \log(q))& O(n^4d\cdot \log(q)) \\\hline
    \textbf{Internal}
    & O(n^4dN + n^2d\cdot \mathsf{MM}(n)+ n^3d^2 )
    & O(n^4dN + n\mathsf{MM}(n,nd))
    & O(n^4dN + n\cdot\mathsf{MM}(n,nd)+\mathsf{H}(nd))\\ \hline
    \end{array}
$$
}